\pdfoutput=1
\documentclass[10pt,final,conference,letterpaper]{IEEEtran}
\usepackage{fancyhdr}

\usepackage[numbers]{natbib}

\usepackage{tikz}
\usepackage{amsmath,amssymb,amsfonts}
\usepackage{subfig}
\usepackage{url}
\usepackage{multirow}
\usepackage{xspace}
\usepackage[inline]{enumitem}

\usepackage[hidelinks]{hyperref}

\usepackage{inconsolata}
\usepackage[frozencache]{minted}

\newcommand{\spack}{Spack\xspace}
\newcommand{\clingo}{\texttt{clingo}}
\newcommand{\perfsubfigwidth}{0.22}
\definecolor{bg}{rgb}{0.95,0.95,0.95}

\begin{document}

\title{Using Answer Set Programming for HPC Dependency Solving}


\author{%
  \IEEEauthorblockN{%
    Todd Gamblin\IEEEauthorrefmark{1}, %
    Massimiliano Culpo\IEEEauthorrefmark{2}, %
    Gregory Becker\IEEEauthorrefmark{1}, %
    Sergei Shudler\IEEEauthorrefmark{1}}
  \IEEEauthorblockA{%
    \IEEEauthorrefmark{1}Lawrence Livermore National Laboratory, Livermore, CA, USA\\
    \{\tt%
    \href{mailto:tgamblin@llnl.gov}{tgamblin}, %
    \href{mailto:becker33@llnl.gov}{becker33}, %
    \href{mailto:shudler1@llnl.gov}{shudler1}%
    \}@llnl.gov}
\IEEEauthorblockA{\IEEEauthorrefmark{2}np-complete, S.r.l., Italy\\
  {\tt\href{mailto:massimiliano.culpo@googlemail.com}{massimiliano.culpo@googlemail.com}}}
}

\maketitle
\thispagestyle{fancy}
\lhead{}
\rhead{}
\chead{}
\lfoot{\footnotesize{
    Preprint, \today.
  }
} \rfoot{}
\cfoot{}
\renewcommand{\headrulewidth}{0pt} \renewcommand{\footrulewidth}{0pt}

\begin{abstract}
  Modern scientific software stacks have become extremely complex, using many
  programming models and libraries to exploit a growing variety of GPUs and
  accelerators. Package managers can mitigate this complexity using dependency solvers,
  but they are reaching their limits. Finding compatible dependency versions is
  NP-complete, and modeling the semantics of package compatibility modulo build-time
  options, GPU runtimes, flags, and other parameters is extremely difficult. Within this
  enormous configuration space, defining a ``good'' configuration is daunting.

  We tackle this problem using Answer Set Programming (ASP), a declarative model for
  combinatorial search problems. We show, using the Spack package manager, that ASP
  programs can concisely express the compatibility rules of HPC software stacks and
  provide strong quality-of-solution guarantees. Using ASP, we can mix new builds with
  preinstalled binaries, and solver performance is acceptable even when considering tens
  of thousands of packages.
\end{abstract}


\section{Introduction}
\label{sec:intro}


Managing dependencies for scientific software is notoriously
difficult~\cite{hoste+:pyhpc12,gamblin+:sc15,dubois2003johnny,hochstein+:2011-build}.
Scientific software is typically built from source to achieve good performance, and
configuring build systems, dependency versions, and compilers requires painstaking care.
In the past decade, the modularity of HPC software has increased due to well known
benefits such as separation of concerns, encapsulation, and code reuse. These enable
application codes to leverage fast math libraries~\cite{hypre,mfem,petsc,trilinos} and
GPU performance portability frameworks like RAJA~\cite{raja} and Kokkos~\cite{kokkos}.
However, the cost of modularity is integration complexity: consumers of components must
ensure that versions and other parameters are chosen correctly to ensure that all
integrated components work together.

\begin{figure*}
  \centering
  \includegraphics[width=\textwidth]{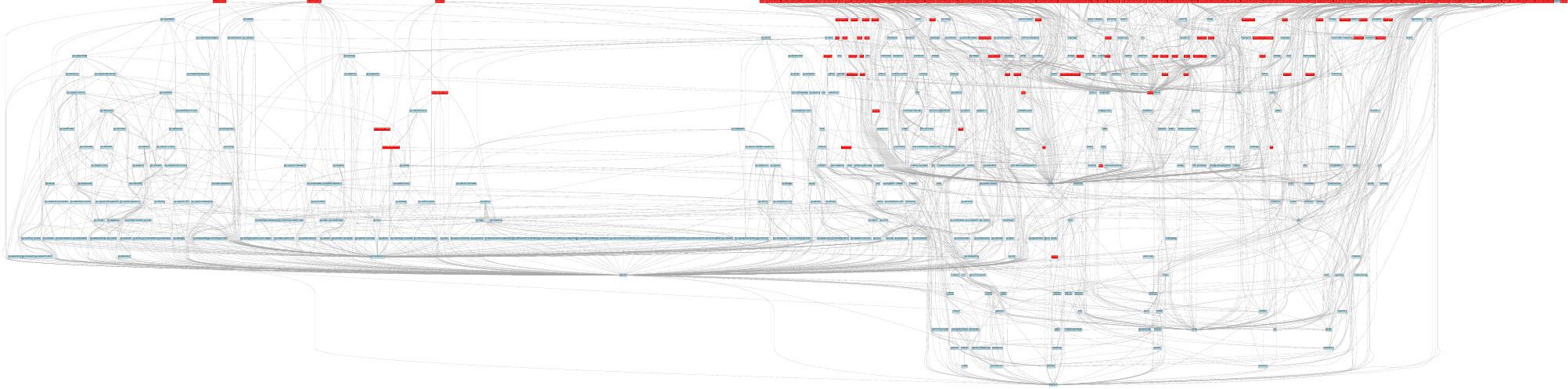}
  \caption{
    Graph of package dependencies in the Extreme Scale Scientific Software Stack (E4S)~\cite{e4s}.
    Around 100 core software products are shown in red, and the 500 required dependencies
    are blue. The precise number of packages required for a deployment of E4S
    is system-specific, depending on platform and hardware (e.g., NVIDIA, AMD GPUs).
    \label{fig:e4s-graph}
  }
\end{figure*}

{\it Package managers} emerged in the mid-1990's to mitigate the complexity of
integrating software packages in Linux distributions. In the past decade, their use has
exploded, particularly within language ecosystems like Python~\cite{pip},
Javascript~\cite{npm}, and Rust~\cite{cargo}, but also within the HPC community. The
HPC-centric package managers Spack~\cite{gamblin+:sc15} and
EasyBuild~\cite{hoste+:pyhpc12} are now widely used for software deployment at HPC
centers, and by developers and users installing their own software. The U.S. Exascale
Computing Project (ECP) has adopted Spack as the distribution tool for its
E4S~\cite{e4s} software stack (shown in Figure~\ref{fig:e4s-graph}), which contains
around 100 core software products and 500 required dependency packages.

Package managers address the integration problem in two main ways. Systems like
EasyBuild and Nix~\cite{dolstra+:icfp08,dolstra+:lisa04} rely on human effort, where
maintainers develop fixed package configurations that define a common stack to be
shared among users. To deviate from the stack, users must update all configuration files
to ensure that versions and other parameters are consistent. More commonly, package
managers like Spack provide flexibility for the end user by incorporating {\it
  dependency solvers} at their core. Users can request arbitrary versions, and packages
declare {\it constraints} that bound the space of compatible configurations. The solver
selects a set of versions and configuration parameters that satisfy the user's
requirements and package constraints.

Dependency solving is NP-complete, even for a ``simple'' configuration space with only
packages and versions~\cite{dicosmo:edos,cox:version-sat}. Here, we focus on Spack's
dependency solver, known as the {\it concretizer}, which adds build options (variants),
compilers, target microarchitectures, and dependency versions to make the space even
larger. Most package managers use their own ad-hoc solver~\cite{abate2020dependency},
and Spack is no different; it has historically used its own {\it greedy} algorithm.
Heuristics were sufficient when there were only 245 packages~\cite{gamblin+:sc15}, but
now, Spack's mainline repository contains over 6,000 packages, each with many
constraints and options, and the original concretizer has begun to show its age. In
particular, it lacks:
\begin{itemize}
\item {\it Completeness}: in a growing number of cases, it may not find a solution even
  though one exists; and
\item {\it Optimality}: it provides no guarantees that it has found the ``best''
  of all valid solutions according to any criteria.
\end{itemize}

An increasing number of package managers use Boolean satisfiability (SAT) solvers to
resolve dependency constraints~\cite{abate2020dependency}, which works well for version
solving with a single optimization objective. With its added dimensions, Spack's solver
encodes a much larger configuration space {\it and} requires multi-objective
optimization. Encoding these semantics in pure SAT is {\it extremely} complex. Instead,
we have turned to Answer Set Programming
(ASP)~\cite{gebser+:asp-book,marek+:asp-origins}, a declarative model that allows us to
encode dependency semantics in a first-order, Prolog-like syntax. ASP solvers reduce
first-order logic programs to SAT and optimization, and they guarantee both complete
and optimal solutions. Unlike Prolog, they are also guaranteed to terminate. ASP
encodings are non-trivial, and this paper presents the first dependency solver capable
of making strong guarantees for the full generality of HPC dependency semantics.
Specifically, our contributions are:

\begin{enumerate}
\item A general mapping of Spack's DSL, compatibility semantics, and optimization rules
  to ASP;
\item A technique to optimize for reuse of existing builds in combinatorial package
  solves;
\item An ASP solver implementation in Spack that enables {\it complete} and {\it
  optimal} solutions; and
\item An evaluation of our system's performance on the E4S repository with tens of
  thousands of packages.
\end{enumerate}

Together, these contributions allow us to replace Spack's original concretizer with a
complete, optimizing solver, written in around 800 lines of declarative ASP. This
represents a leap forward in capability, maintainability, and extensibility to the
ever-increasing complexity of HPC software.

\section{Package Managers}
\label{sec:package-managers}

The idea of ``software release management'' for integrated sets of packages dates back to
a 1997 paper from Van der Hoek et al.~\cite{van1997software}, and the first Linux
package managers emerged around the same time~\cite{rpm,apt}. Since then, the number and
complexity of software packages has grown enormously, as have the use cases for package
managers.

\subsection{Single-prefix Package Managers}

Package managers such as RPM, Yum, and APT~\cite{foster+:rpm03,silva:apt01,yum} are used
to manage the system packages of most Linux distributions. These tools focus on managing
one software stack, built with one compiler, which works well for system software and
drivers that underlie all software on the system. They are designed so that software can
be upgraded in place for bug fixes and security updates. All software goes in to a single
{\it prefix}, e.g., {\tt /usr}, and for this reason, only one version of any package can
be installed at a time. Upgrades are prioritized over reproducibility, and users do not
have great flexibility over precise package versions and configurations.

Even in this limited package/version configuration space, dependency solving is
NP-complete~\cite{dicosmo:edos,mancinelli+:ase06-foss-distros} because any two
non-overlapping version constraints can cause a conflict. The Common Upgradeability
Description Format (CUDF) attempts to standardize an interface for external package
solvers~\cite{abate2012dependency,abate-2013-modular-package-manager} and has enabled
many encodings of the single-prefix upgrade problem: Mixed-Integer Linear Programming,
Boolean Optimization, and Answer Set
Programming~\cite{tucker+:icse07-opium,michel+:lococo2010,argelich+:lococo2010,gebser+:2011-aspcud}.
Most modern package solvers still use ad-hoc implementations instead of a common
external solver~\cite{abate2020dependency}.

\subsection{Language Package Managers}

Most modern language implementations also include their own package
managers~\cite{npm,pip,cargo,weizenbaum:pubgrub18}. The solver requirements for language
package managers are different from those in a Linux distribution; they typically allow
multiple versions of the same package to be installed to support developer workflows and
reproducibility. Conflicts can arise, as most language module systems do not support
{\it linking} multiple versions of one package into a single executable. Javascript is a
notable exception to this. Language package managers still solve only for package
names and versions, ignoring compilers, build options, targets and other options that
expand the solution space. Most language package managers implement their own ad-hoc
solvers. Historically, these solvers were not complete, as
implementing a performant SAT solver in all but the lowest-level languages is quite
difficult. However, there has been a recent trend towards complete
solvers~\cite{pip-new-resolver,weizenbaum:pubgrub18} as complexity of language
ecosystems has grown. These systems do not support optimization beyond selecting recent
versions~\cite{abate2020dependency}.

\subsection{Functional and HPC Package Managers}

So-called ``functional'' package managers like
Nix~\cite{dolstra+:icfp08,dolstra+:lisa04} and Guix~\cite{courtes-guix-2015} allow users
to install arbitrarily many configurations of a given package. Like many HPC
deployments, they install each configuration to a unique prefix, but instead of a
human-readable name, they ensure that each installation gets a unique prefix by using a
hash of the bits of the installation itself. These systems do not use solvers to resolve
dependencies. Rather, they have very little conditional logic and rely on the specific
package recipes in a repository. As mentioned, EasyBuild~\cite{hoste+:pyhpc12} is
similar in this regard, as it allows multiple configurations of packages by duplicating
package recipes and adhering to a strict naming convention to differentiate different
installations. In all of these systems, users must modify code in package recipes to
change the way dependencies are resolved.

\section{Spack}
\label{sec:software-model}

\spack~\cite{gamblin+:sc15} is a package manager designed to support High Performance
Computing (HPC). Like functional package managers, \spack installs packages in separate
prefixes to allow arbitrarily many installations to coexist. This enables users to build
many different variants of a package, e.g., with different compilers, different MPI
implementations, different flags, or with combinations of all three of these parameters.
This greatly expands the space of possible build configurations.

At a high level, \spack can be broken down into three primary components. The {\it spec
  syntax} allows users to easily specify and constrain builds on the command line. The
{\it package domain-specific language (DSL)} expresses parameterized build recipes for
software packages, and the {\it concretizer} is Spack's dependency solver. The
concretizer combines {\it abstract} (i.e., underspecified) spec constraints from the
user and from packages, and it produces a {\it concrete}, or fully specified package
spec that can be installed. We provide an overview of these pieces here but full details
are in the original paper~\cite{gamblin+:sc15}.

\subsection{Spec Syntax}
\label{sec:specs}

Spack calls its internal dependency graphs ``specs'' because they specify parameters of
a software installation along with those of its dependencies. Specs are directed acyclic
graphs where nodes are packages and edges are dependency relationships. The ``spec
syntax'' is a shorthand for expressing constraints on these graphs. Each node of the
graph has parameters, including
\begin{enumerate*}
\item the package name;
\item the version to build;
\item variants (compile-time build options);
\item the compiler to build with and its version;
\item compiler flags;
\item the target operating system;
\item the target microarchitecture; and
\item the package installation's dependencies.
\end{enumerate*}

\begin{table}
  \centering
  \footnotesize
  \begin{tabular}{|c|l|p{3.5cm}|}
    \hline
    \textbf{Sigil} & \textbf{Examples} & \textbf{Meaning} \\
    \hline\hline
    \texttt{\%} & \texttt{hdf5\%gcc}    & Use a particular compiler \\
    \hline
    \texttt{@}  & \texttt{hdf5@1.10.2}  & Require version(s) \\
    & \texttt{hdf5\%gcc@10.3.1} & Require compiler version(s) \\
    \hline
    \texttt{+}  & \texttt{hdf5+mpi}     & \multirow{2}{3.5cm}{Enable (\texttt{+})/disable (\texttt{\~{}}) variant} \\
    \texttt{\~} & \texttt{hdf5\~{}mpi} &    \\
    \hline
    \texttt{key=value} & \texttt{hdf5 mpi=true} & \multirow{3}{3.5cm}{Require a particular variant or build target value} \\
    & \texttt{hdf5 api=default} &      \\
    & \texttt{hdf5 target=skylake} &  \\
    \hline
  \end{tabular}
  \caption[Table caption text]{
    Spec sigils in Spack
    \label{tab:sigil}
    \vspace{-1em}
  }
\end{table}

The spec syntax in \spack allows users to specify preferences within this combinatorial
build space. In the simplest case, a user might want to build a particular package {\it
  without} any concern for its configuration. In this case they would write:
\begin{minted}[fontsize=\small,bgcolor=bg]{text}
spack install hdf5
\end{minted}
Spack will install {\tt hdf5} without any particular customization.
Table~\ref{tab:sigil} shows examples of {\it sigils} that can be used to specify
specific constraints on specs, including any of the attributes listed above. The spec
syntax is fully recursive, in that users can specify constraints on dependencies using
the {\tt \^{}} (``depends on'') sigil. For example, the spec below:
\begin{minted}[fontsize=\small,bgcolor=bg]{text}
hdf5@1.10.2 ^zlib%gcc ^cmake target=aarch64
\end{minted}
means ``any possible build of the package \texttt{hdf5} at version {\tt 1.10.2} that
depends on a build of the package \texttt{zlib} built with the \texttt{gcc} compiler and
a build of the package \texttt{cmake} built to target the \texttt{aarch64}
architecture''. While users could completely constrain a build via the spec syntax, in
practice they only care about a comparatively small set of constraints, and they rely on
concretization to select values for unspecified parameters. We refer to a portion of the
build space in \spack as an ``abstract spec'', and a fully specified build as a
``concrete spec''.

\subsection{Package DSL}
\label{sec:package-dsl}

\begin{figure}
\begin{minted}[linenos,numbersep=-5pt,fontsize=\footnotesize]{python}
  # This is the class name for the package `example`
  class Example(Package):
      """Example depends on zlib, mpi, and optionally bzip2"""

      version("1.1.0")
      version("1.0.0")

      variant("bzip", default=True, description="enable bzip")

      # Depends on bzip2 or later when bzip is enabled
      depends_on("bzip2@1.0.7:", when="+bzip")
      depends_on("zlib")  # depends on zlib
      # Newer versions require newer versions of zlib
      depends_on("zlib@1.2.8:", when="@1.1.0:")

      # Depends on *some* MPI implementation
      depends_on("mpi")

      # Known failure when building with intel compilers
      conflicts("%intel")
      # Does not support architectures derived from ARM64
      conflicts("target=aarch64:")

      def install(self, spec, prefix):
          # Translate spec into arguments to build system
          config_args = ['--with-zlib=%s' % spec['zlib'].prefix]

          if 'bzip2' in spec:
              bz_arg = '--with-bzip=%s' % spec['bzip2'].prefix
          else:
              bz_arg = '--without-bzip'
          config_args.append(bz_arg)

          # Run the build
          configure(*config_args)
          make()
          make('install')
\end{minted}
\caption{
  A {\tt package.py} file written in Spack's embedded DSL.
  \label{fig:example-spack-package}
  \vspace{-3em}
}
\end{figure}

Figure~\ref{fig:example-spack-package} shows Spack's package DSL. Unlike many systems, a
package in \spack is a {\it parameterized} Python class. There is {\it one} package
recipe for each package, and it is {\it instantiated} for each concrete spec that Spack
builds. The {\tt install()} function on line 24 takes a fully concrete spec as one of
its parameters, and its job is to install {\it that} particular configuration of the
{\tt example} package into the specified {\tt prefix}.

The install instructions on lines 24-37 do the work of the build, but more interesting
for this paper are the metadata directives on lines 5-22. This package has two possible
versions, {\tt 1.0.0} and {\tt 1.1.0}. It has one build option, or ``variant'', called
{\tt bzip} that enables the dependency on {\tt bzip2}. Its dependency on {\tt zlib} is
dependent on its version. Conditions like these are specified using {\tt when} clauses
within directives. It has some {\tt conflicts}---conditions under which it is known not to build
successfully. Dependency constraints, conflicts, and {\tt when} clauses are all
specified concisely using the  spec syntax.
The dependency on {\tt mpi} on line 17 is special. Spack provides support for packages
that are source-compatible (API-compatible), like MPI. There is no concrete {\tt mpi}
package in Spack, rather there are several {\tt mpi} {\it providers}, like {\tt mpich},
{\tt openmpi}, etc. This package can be instantiated with {\it any} of these MPI
providers. We thus call {\tt mpi} a``virtual dependency''. {\tt blas} and {\tt lapack}
are two other common examples in HPC.

\subsection{Concretization}

Spack can build the {\tt example} package in Figure~\ref{fig:example-spack-package} in
{\it many} different ways. The metadata in the package DSL exposes a combinatorial build
configuration space, and it is the {\it concretizer}'s job to select a configuration.
The {\it concretizer} is Spack's dependency solver, so called because it converts
partially constrained \emph{abstract specs} to fully constrained \emph{concrete specs}
that can be built when combined with installation methods in the package DSL.
There are three primary inputs to concretization in \spack: specs on the command line,
constraints from the package DSL, and additional preferences from user configuration
files. Without loss of generality, we consider only the first two here.
%
%
%
\subsubsection{Validity}

On a conceptual level, we can think of the concretizer as accomplishing two tasks.
First, it determines the set of packages needed in the spec DAG. Second, it fills in any
missing nodal parameters (enumerated in Section~\ref{sec:specs}). Every decision made by
the concretizer must consider constraints from all the input sources; any unsatisfied
constraints imply an {\it invalid} solution. We say a solution is {\it valid} if and only if:
\begin{itemize}
\item all virtuals are replaced with concrete dependencies;
\item all dependencies are resolved;
\item all node parameters have been assigned values; and
\item all input constraints are satisfied.
\end{itemize}

\subsubsection{Completeness}
The original Spack concretizer used a greedy, fixed-point algorithm. The issue with this
algorithm (and all greedy algorithms) is that it only made {\it local} decisions about
dependencies and node parameters, and it could not backtrack to undo decisions that it
made. A solver is {\it complete} if it finds a solution when one exists; this property
of the original concretizer made it {\it incomplete}.

Consider our package \texttt{example} from Section~\ref{sec:package-dsl}, we will
concretize the abstract spec:
\begin{minted}[fontsize=\small,bgcolor=bg]{text}
example@1.0.0 ^zlib@1.2.11
\end{minted}
We will assume for the moment that neither \texttt{zlib} nor \texttt{bzip2} have any
additional dependencies. The solver needs to consider contraints from package files
and from the abstract spec supplied on the command line. Here, those constraints are:

\begin{itemize}
\item {\tt example@1.0.0} from the abstract spec: ``example v1.0.0''
\item {\tt bzip2@1.0.7:} from {\tt example}: ``bzip2 v1.0.7 or higher''
\item {\tt zlib@1.2.11} from abstract spec: ``zlib version 1.2.11''
\item {\tt mpi} from {\tt example}: ``any MPI implementation''
\end{itemize}

The original concretizer would simply proceed until all decisions had been made,
stopping at the first conflict it found with any prior decision. The
resulting concrete spec might be:
\begin{minted}[fontsize=\footnotesize,bgcolor=bg]{text}
example@1.0.0+bzip%gcc@11.2.0 arch=linux-centos8-skylake
  ^bzip2@1.0.8+pic%gcc@11.2.0 arch=linux-centos8-skylake
  ^zlib@1.2.11+pic%gcc@11.2.0 arch=linux-centos8-skylake
  ^mpich@3.1 pmi=pmix %gcc@11.2.0 arch=linux-centos8-skylake
\end{minted}
Here, all dependencies are expanded, all virtual packages have been replaced, all node
parameters (variants, compilers, targets, etc.) have been assigned and no constraints
are violated. Decisions regarding node attributes not specified in the abstract spec nor
in the package constraints were made from defaults. But, imagine that {\tt mpich} had a
conflict with {\tt bzip2@1.0.7}. In our example, we were lucky that the concretizer
selected {\tt bzip2@1.0.8} to satisfy the constraint {\tt bzip2@1.0.7:} (recall that
means \texttt{bzip2} at version 1.0.7 or higher). Had the default choice been {\tt
  bzip2@1.0.7}, there would have been no way for the algorithm to backtrack, and it would
fail with no solution.

\subsubsection{Optimality}
There are usually many {\it valid} solutions to concretization problems, but users
prefer some solutions over others. For a fresh installation, users generally prefer the
newest version of a package. They prefer certain defaults for build options and
variants, and HPC users tend to prefer specific microarchitecture targets with vector
instructions, e.g., {\tt skylake} or {\tt cascadelake} with AVX-512 support vs. generic
{\tt x86\_64}. We say a solver is {\it optimal} when it can find the best valid solution
according to some set of formal criteria. Since the original concretizer was not
complete, it also could not be optimal, as it could never consider all possibilities.
Note that this definition of optimality does {\it not} refer to the performance of built
binaries, though it may be related depending on the criteria chosen (we discuss this
in later sections).

In addition to these issues, the original concretizer was implemented directly on the
dependency graph structure in memory. Any new semantics had to be implemented with
complex graph algorithms, and graph nodes and edges had to be configured and updated
manually. This process was error-prone and limited the speed of development and the
complexity of the semantics that could be considered.

\section{Answer Set Programming}
\label{sec:asp}

The concretization problem is essentially a combinatorial search problem. While many
package managers use homegrown SAT solvers for this type of problem, encoding the
complexity of our domain in pure Boolean SAT would be extremely tedious. To manage the
complexity, we leverage Answer Set Programming (ASP), a form of declarative programming
that allows us to formally specify our semantics in first-order logic with variables and
quantification, in addition to Boolean operators. The input language for ASP is similar
to Prolog~\cite{baral_2003}. Unlike Prolog, ASP has no operational semantics and is not
Turing-complete. Rather, ASP converts first-order logic programs (with quantifiers and
variables) to {\it propositional} programs (with no variables). It then uses techniques
borrowed from SAT solvers to find solutions. One major benefit of this approach is that
unlike Prolog, ASP programs are guaranteed to terminate.

In the remainder of this section, we illustrate how ASP can simplify the specification
and maintenance of our concretization algorithm while also affording strong guarantees
than we can implement ourselves with simple heuristics.

\subsection{ASP Syntax}

ASP programs are comprised of ``terms'', which can be Boolean atoms or a functions whose
arguments may also be terms. A term followed by a period ({\tt .}) is called a ``fact''.
The following is a simple program comprised entirely of facts:
\begin{minted}[fontsize=\small, bgcolor=bg]{prolog}
  optimize_for_reuse.
  node("hdf5").
  depends_on("hdf5", "mpi").
\end{minted}
Note that these functions are not imperative; rather this snippet can be ready roughly
as ``{\tt optimize\_for\_reuse} is enabled. A node called {\tt hdf5} exists. {\tt hdf5} depends on {\tt mpi}.''

In addition to facts, ASP programs contain rules, which can derive additional facts. An
ASP rule has a {\it head} and a {\it body}, separated by \texttt{:-}. The \texttt{:-}
can be read as ``if'' --- the head (left side) is true {\it if} the body (right side) is true.
Terms in the body of a rule can also be preceded by the keyword ``not'' to imply the
head based on their negation. Logical ``and'' is represented by a comma in the rule
body, and ``or'' is represented by repeating the head with a different rule body.

ASP programs can contain variables, represented by capitalized words. Variables are
scoped to the rule or fact in which they appear; the variable {\tt Package} may be used
in unrelated rules without any scoping issues. Rules are instantiated with all possible
substitutions for variables. For example:
\begin{minted}[fontsize=\small, bgcolor=bg]{prolog}
  node("hdf5").
  depends_on("hdf5", "mpi").
  node(Dependency) :-
      node(Package), depends_on(Package, Dependency).
\end{minted}
The rule here translates roughly to ``If a package is in the graph and it depends on
another package, that package must be in the graph'', and the rule in the program above
will derive the fact {\tt node("mpi")} when it is instantiated with {\tt
  node("hdf5")} and {\tt depends\_on("hdf5", "mpi")}. This is the essence of how
  we model dependencies in ASP.

\subsection{Integrity Constraints and Choices}

Two additional types of rules are important to understand the power of ASP:
\textit{integrity constraints} and \textit{choice rules}.
Integrity constraints allow the programmer to rule out swathes of the answer set space
by specifying conditions not to allow. In ASP, these are represented by a headless rule:
\begin{minted}[fontsize=\small, bgcolor=bg]{prolog}
  node("hdf5").
  depends_on("hdf5", "mpi").
  node(Dependency) :-
      node(Package), depends_on(Package, Dependency).
  :- depends_on(Package, Package).
\end{minted}
This constraint says ``a package cannot depend on itself''.

\subsection{Choice Rules}

Choice rules give the solver freedom to choose from possible options. A choice rule is
(optionally) constrained on either side to specify the minimum (left) and maximum
(right) number of choices the solver can select.
\begin{minted}[fontsize=\small, bgcolor=bg]{prolog}
  node("hdf5").
  depends_on("hdf5", "mpi").
  node(Dependency) :-
      node(Package), depends_on(Package, Dependency).
  :- depends_on(Package, Package).

  possible_version("hdf5", "1.13.1").
  possible_version("hdf5", "1.12.1").
  1 { version(P, V) : possible_version(P, V) } 1 :-
      node(P).
\end{minted}
The choice rule on the last line means ``If a package is in the graph, assign it
exactly one of its possible versions.'' The $\{a:b\}$ syntax here has the same meaning
as in formal set theory; it denotes the set of all $a$ such that $b$ is true. The
numerals on either side of the choice rule are {\it cardinality constraints}; they specify
lower and upper bounds on the size of the set. Choice rules are similar to guessing;
they give the solver options for paths to explore next.


\subsection{Optimization}

In addition to constraints on the solution space, ASP allows optimization criteria.
Optimization criteria allow the solver to return a single answer set from the among all
valid answer sets, and they guarantee that the chosen answer set minimizes or maximizes one
or many criteria. There are often many valid solutions to a dependency resolution query,
but only the optimal solution is relevant once we find it.
\begin{minted}[fontsize=\small, bgcolor=bg]{prolog}
  node("hdf5").
  depends_on("hdf5", "mpi").
  node(Dependency) :-
      node(Package), depends_on(Package, Dependency).
  :- depends_on(Package, Package).

  possible_version("hdf5", "1.13.1", 0).
  possible_version("hdf5", "1.12.1", 1).

  1 { version(P, V) : possible_version(P, V) } 1 :-
      node(P).

  version_weight(P, V, Weight) :-
      possible_version(P, V, Weight).
  #minimize{ W@3,P,V : version_weight(P, V, W)}.
\end{minted}
This optimization constraint says ``I prefer (at priority 3) solutions that minimize the
sum of the weights ({\tt W}) of the package ({\tt P}) versions ({\tt V}), for all
packages and versions.'' A detailed discussion of the syntax for optimization is beyond
the scope of this paper; it suffices to know that multi-objective optimization across a
variety of criteria is possible in ASP.

This small and very incomplete program already shows the power of ASP. In a SAT
solver, the programmer would need to write out explicitly the rule for deriving
dependency nodes from dependent nodes for every possible pair of packages. Even worse,
our choice rule for selecting versions would need to be written out for every
combination of package and version, along with a combinatorial set of clauses
ensuring that for any versions \texttt{$a_1, ..., a_n$}, exactly one
\texttt{$a_i$} is set but not any \texttt{$a_j$} for all $i\neq{j}$. ASP
makes this much simpler.


\subsection{Grounding and Solving}

ASP solvers search the input program for {\it stable models}, also known as
\textit{answer sets}. An answer set is a set of true atoms for which every rule in the
input program is idempotent. It is similar to a fixed point, or the solution to a system
of (Boolean) equations. When ASP first emerged in the late 1990's, it used custom solvers.
However, the problem domain of ASP is NP-complete, and the 2000's brought great
advances~\cite{moskewicz2001chaff} in solving the canonical NP-complete problem: Boolean
satisfiability (SAT). Modern ASP solvers exploit high-performance techniques developed
in the SAT community~\cite{gebser+:asp-book}.

\begin{figure*}[t]
  \centering
  \includegraphics[width=.8\textwidth]{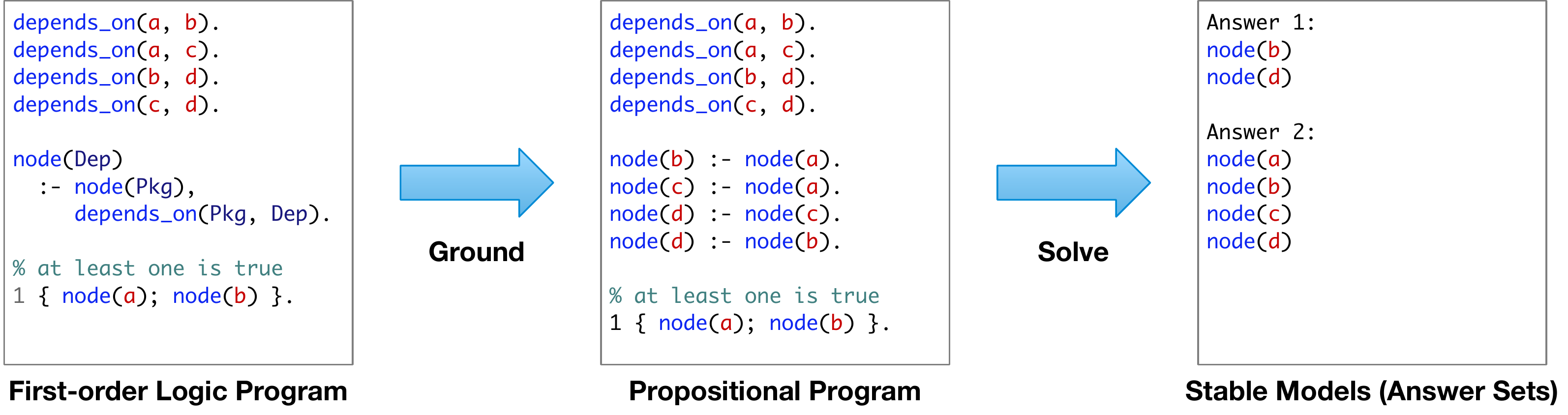}
  \caption{
    Grounding and solving in Answer Set Programming.
    \label{fig:ground-solve}
    \vspace{-1em}
  }
\end{figure*}

The ASP input language, with first-order rules and variables, is very expressive, but
SAT solvers deal with {\it propositional} logic programs that have no variables. A logic
expression that has no variables is called a {\it ground} term, and {\it grounding} is
the process of converting a first-order ASP rules to a set of ground rules. We call
these the ``ground instances'' of the rule.

Figure~\ref{fig:ground-solve} shows the grounding and solving process at a very high
level. On the left is an ASP program with four facts, one first-order rule, and a choice
rule that says we must choose at least one of {\tt node(a)} or {\tt node(b)} to be in
the solution. Grounding this program {\it instantiates} the first-order rule with all
possible ground atoms that can be substituted into its body. The ground instances are
built from input {\tt depends\_on()} facts, {\tt node(a)} and {\tt node(b)} from the
choice rule, as well as {\tt node(c)} and {\tt node(d)}, which appear in the heads of
ground rules instantiated from {\tt node(a)} and {\tt node(b)}. Note that the ground
instances are simplified---they omit {\tt depends\_on()} terms because these are facts
in the input, and therefore always true. Grounders perform many such optimizations to
prune the propositional program and to make the solver more efficient.

Propositional logic programs can be reduced to SAT and solved using techniques similar
to those used in modern SAT solvers~\cite{gebser+:asp-book}. For this program, there are
two stable models: one with only {\tt node(b)} and {\tt node(d)} (when only {\tt
  node(b)} is selected by the choice rule) and one with all four nodes (when only {\tt
  node(a)} or both {\tt node(a)} and {\tt node(b)} are selected). It is easy to see from
this result how we could read in these lists of nodes and construct graphs from them.
This how Spack reads solutions in from the solver.

We use the popular {\tt clingo}~\cite{gebser+:aicomm11} system, which includes a
grounder ({\tt gringo}) and a solver ({\tt clasp}). The search algorithm used in {\tt
  clasp} traces its roots to the well known Davis–Putnam–Logemann–Loveland (DPLL)
algorithm~\cite{dp-sat,dpll-sat}, but uses modern extensions like Conflict-Driven Clause
Learning (CDCL) for high performance~\cite{moskewicz2001chaff}. {\tt clasp} can also
do MaxSAT-style optimization.
%
The internals of {\tt clingo} are beyond the scope of this paper, but is important to
understand that because it effectively performs an exhaustive combinatorial search, it
guarantees completeness and optimality. There are no inputs for which {\tt clingo} will
return a false negative (claim compatible rules are incompatible) and solutions
are guaranteed to be optimal according to provided criteria.

\section{Modeling Software Dependencies with ASP}
\label{sec:asp-model}

At a high-level, using \clingo, our concretizer combines:
\begin{enumerate}
\item Many \emph{facts} characterizing the problem instance;
\item A small \emph{logic program} encoding the rules and constraints of the software model; and
\item Optimization criteria that define an {\it optimal} model.
\end{enumerate}
The facts are always generated starting from one or more \emph{root specs}, and they
account for metadata from all package recipes of possible dependencies, as well as
the current state of \spack{} in terms of configuration and installed software. This
directive:
\begin{minted}[fontsize=\small, bgcolor=bg]{python}
version('1.2.11', sha256='ah45rstgef...')
\end{minted}
in \texttt{zlib}'s recipe is translated to the following fact:
\begin{minted}[fontsize=\small, bgcolor=bg]{prolog}
version_declared("zlib", "1.2.11", 0).
\end{minted}
where {\tt 0} is the preference weight of this version. Similarly:
\begin{minted}[fontsize=\small, bgcolor=bg]{prolog}
zlib@1.2.11
\end{minted}
generates the following three facts:
\begin{minted}[fontsize=\small, bgcolor=bg]{prolog}
root("zlib").
node("zlib").
version_satisfies("zlib","1.2.11").
\end{minted}
stating that \texttt{zlib} is a root node and should satisfy a version requirement. A
typical solve has around $10k-100k$ facts that encode dependencies, variants,
preferences, etc.

The logic program encodes the software model used by \spack{} and only contains
first-order rules, integrity constraints, and optimization objectives. The declarative
nature of ASP makes it easy to enforce certain properties on the solution. For
instance, these three lines ensure that we never have a cyclic dependency in a DAG:
\begin{minted}[fontsize=\small, bgcolor=bg]{prolog}
path(A, B) :- depends_on(A, B).
path(A, C) :- path(A, B), depends_on(B, C).
:- path(A, B), path(B, A).
\end{minted}
The first rule is the base case: if {\tt A} depends on {\tt B} there is a path from {\tt
  A} to {\tt B}. The second rule defines paths to transitive dependencies recursively:
if there is a path from {\tt A} to {\tt B} and {\tt B} depends on {\tt C}, there is a
path from {\tt A} to {\tt C}. The final line is an \emph{integrity constraint} banning
paths from {\tt A} to {\tt B} and paths from {\tt B} to {\tt A} from occurring together
in a solution.

To give a rough idea of the compactness and expressiveness of the ASP encoding, the
entire logic program for the software model described here is around $800$ lines. The
concretization process is straightforward to follow conceptually within \spack:
\begin{enumerate}
\item Generate facts for all possible dependencies/installs \footnotemark;
\item Send logic program and facts to the solver;
\item Retrieve the best stable model; and
\item Build an {\it optimal} concrete DAG from the model.
\end{enumerate}
\footnotetext{We stress that the logic program changes only when the underlying software
  model changes, as opposed to the generated facts that are different whenever the root
  spec to be concretized or \spack's configuration changes.} The word ``optimal'' is
emphasized since, while rules and integrity constraints fully determine if a solution is
valid, we need optimization targets to select one of the many possible solutions in a
way that fits user's expectations.

A good example to illustrate this point is target selection for DAG nodes. In \spack{}
each node being built has a target microarchitecture associated with it, and we want to
use the best target possible while respecting the constraints coming from the compiler
(for example, {\tt gcc@4.8.3} cannot generate optimized instructions for {\tt skylake}
processors). Previously this required some complicated logic mixed with the rest of the
solve. The introduction of \clingo{} greatly simplified the problem definition. A
cardinality constraint is used to express that the solver must choose one and only one
target per node:
\begin{minted}[fontsize=\small, bgcolor=bg]{prolog}
1 { node_target(Package, Target) : target(Target) } 1
    :- node(Package).
\end{minted}
A user's choice will force the target of a node:
\begin{minted}[fontsize=\small, bgcolor=bg]{prolog}
node_target(P, T) :- node(P), node_target_set(P, T).
\end{minted}
An integrity constraint prevents choosing targets not supported by the chosen compiler:
\begin{minted}[fontsize=\small, bgcolor=bg]{prolog}
:- node_target(P, T),
   not compiler_supports_target(C, V, T),
   node_compiler(P, C),
   node_compiler_version(P, C, V).
\end{minted}
These three statements fully describe the characteristics of a valid solution. To pick
the \emph{optimal} solution we also \emph{weight} the possible targets (the lower the
weight, the best the target) and optimize over the sum of target weights:
\begin{minted}[fontsize=\small, bgcolor=bg]{prolog}
node_target_weight(P, W) :-
  node(P), node_target(P, T), target_weight(T, W).
#minimize { W@5,P : node_target_weight(P, W) }.
\end{minted}

\subsection{Generalized Condition Handling}
\label{subsec:generalizedcond}
A unique feature of \spack, as a package manager, is that it optimizes not only for
versions but for many other aspects of the build, e.g., which compiler to use,
which microarchitecture to target, etc. The DSL used for package recipes reflects this
complexity by having multiple directives to declare properties or constraints
on software packages, as seen in Section~\ref{sec:software-model}.

One interesting abstraction that we observed while coding the ASP logic program is
that each of these directives can be seen as a way to impose additional constraints on
the solution, conditional on other constraints being met. For instance, the following
directive in a package:
\begin{minted}[fontsize=\small, bgcolor=bg]{python}
depends_on('hdf5+mpi', when='+mpi')
\end{minted}
means that, if the spec has the {\tt mpi} variant turned on, then it depends on {\tt hdf5+mpi}. Similarly:
\begin{minted}[fontsize=\small, bgcolor=bg]{python}
provides('lapack', when='@12.0:')
\end{minted}
means that a package provides the {\tt LAPACK} API if its version is $12.0$ or greater.
This property allowed to encode all the directives as \emph{generalized conditions}, where most of the semantics is encoded abstractly in a few lines of the logic program.

Getting back to a simple example, the snippet below:
\begin{minted}[fontsize=\small, bgcolor=bg]{python}
class H5utils(AutotoolsPackage):
    depends_on('png@1.6.0:', when='+png')
\end{minted}
is translated to the following facts:
\begin{minted}[fontsize=\scriptsize, bgcolor=bg]{prolog}
condition(153).
condition_requirement(153, "node", "h5utils").
condition_requirement(153, "variant_value", "h5utils", "png", "true").
imposed_constraint(153, "version_satisfies", "libpng", "1.6.0:").
dependency_condition(153, "h5utils", "libpng").
\end{minted}
when setting up the problem to be solved by \clingo. The important points to note are that:

\begin{itemize}
\item Each directive is associated with a unique global ID.
\item Constraints are either ``requirement'' or ``imposed''.
\item Different type of conditions have different semantics\footnotemark
\end{itemize}
\footnotetext{For instance, the {\tt dependency\_condition} fact is present only for {\tt depends\_on} directives and activates logic that is specific to dependencies.}

The code to trigger and impose general conditions in the logic program is surprisingly simple to read. ASP conditional rules allow us to effectively build new rules from input facts:
\begin{minted}[fontsize=\small, bgcolor=bg]{prolog}
condition_holds(ID) :-
 condition(ID);
 attr(N, A1)    : condition_requirement(ID, N, A1);
 attr(N, A1, A2): condition_requirement(ID, N, A1, A2).
\end{minted}
based on their number of arguments, or \emph{arity}. Other rules:
\begin{minted}[fontsize=\small, bgcolor=bg]{prolog}
attr(N, A1)     :- condition_holds(ID),
                   imposed_constraint(ID, N, A1).
attr(N, A1, A2) :- condition_holds(ID),
                   imposed_constraint(ID, N, A1, A2).
\end{minted}
enforce the imposed constraints when a condition holds.

\subsection{Usability Improvements due to \clingo}
As mentioned, the original concretizer was incomplete and could fail to find a solution
when one exists. Users would work around such false negatives by overconstraining
problematic specs, to help the solver find the right answer. This could become very
tedious.


\subsubsection{Conditional Dependencies}
A prominent example of this behavior is with packages having dependencies conditional to a variant being set to a non-default value. Let's take for instance {\tt hpctoolkit}, which has the following directives:

\begin{minted}[fontsize=\small, bgcolor=bg]{python}
class Hpctoolkit(AutotoolsPackage):
    variant('mpi', default=False, description='...')
    depends_on('mpi', when='+mpi')
\end{minted}

Trying to use the old algorithm to concretize {\tt hpctoolkit \^{}mpich} fails like this:

\begin{minted}[fontsize=\footnotesize, bgcolor=bg]{console}
$ spack spec hpctoolkit ^mpich
Input spec
--------------------------------
hpctoolkit
    ^mpich

Concretized
--------------------------------
==> Error: Package hpctoolkit does not depend on mpich
\end{minted}
since the greedy algorithm would set variant values before descending to dependencies.
Since no value is specified for the {\tt mpi} variant, the value chosen is
{\tt false} (the default) which leads to {\tt hpctoolkit} having no dependency on
{\tt mpi}. The workaround required users to understand the conditional dependency and
write {\tt hpctoolkit+mpi \^{}mpich} to concretize successfully. With \clingo{}, the
concretizer simply {\it finds} the correct value for the {\tt mpi} variant, as
setting it is the only way for {\tt mpich} to be part of the solution.



\subsubsection{Conflicts in Packages}
Before using \clingo{}, conflicts in packages were only used to \emph{validate} a
solution computed by the greedy-algorithm. If the solution matched any conflict,
\spack{} would have errored and hinted the user on how to overconstrain the initial spec
to help concretization. With ASP, conflicts are generalized as constraints during the
solve\footnotemark{} and \spack{} no longer needs to ask the user to be more specific.
\footnotetext{With \clingo{} conflicts are treated as shown in
  Section~\ref{subsec:generalizedcond} and, by imposing \emph{integrity constraints} on
  the problem, they effectively prevent portions of the search space from being
  explored}

\subsubsection{Specialization on Providers of Virtual Packages}
Using \clingo{} also enabled more complex use cases e.g. imposing constraints on
specific virtual package providers. A simple example of that is given by the
{\tt berkeleygw} package, which has the following directives:

\begin{minted}[fontsize=\small, bgcolor=bg]{python}
class Berkeleygw(MakefilePackage):
    variant('openmp', default=True)
    depends_on('lapack')
    depends_on('openblas threads=openmp',
               when='+openmp ^openblas')
\end{minted}

The last directive forces {\tt openblas} to have {\tt openmp} support if
{\tt berkeleygw} has {\tt openmp} support and {\tt openblas} has been chosen as
a provider for the mandatory {\tt lapack} virtual dependency. Conditional constraints
of that complexity could not be expressed before, because the solver would select
defaults for {\tt openblas} before evaluating the conditional constraint on it.

\subsection{Optimization Criteria}

\begin{table}[t]
\centering
\begin{tabular}{| c || l |}
 \hline
 Priority & Criterion (to be minimized) \\
 \hline\hline
 1  & Deprecated versions used \\
 \hline
 2  & Version oldness (roots) \\
 3  & Non-default variant values (roots) \\
 4  & Non-preferred providers (roots) \\
 5  & Unused default variant values (roots) \\
 \hline
 6  & Non-default variant values (non-roots) \\
 7  & Non-preferred providers (non-roots) \\
 \hline
 8  & Compiler mismatches \\
 9  & OS mismatches \\
 10 & Non-preferred OS's \\
 \hline
 11 & Version oldness (non-roots) \\
 12 & Unused default variant values (non-roots) \\
 \hline
 13 & Non-preferred compilers \\
 14 & Target mismatches \\
 15 & Non-preferred targets \\
 \hline
\end{tabular}
\caption{
  Spack's optimization criteria in order of priority.
  \label{table:optimization-criteria}
  \vspace{-1em}
}
\end{table}

The conditional logic presented here provides a great deal of flexibility, but with this
flexibility we also need to have sensible defaults for users who do not want to think
about all of the degrees of freedom Spack allows. Coming up with ``intuitive'' solutions
to the package configuration problem is surprisingly difficult, but we have developed
the list of optimization criteria in Figure~\ref{table:optimization-criteria} based on
our experiences interacting with users and facilities. There are currently 15 criteria
we consult at to pick the best valid solution for a package DAG. All are minimization
criteria, and they are evaluated in lexicographic order, i.e., the highest priority
criteria are optimized first, then the next highest, and so on.

Our top priority (1) is to avoid software versions that have been deprecated due to
security concerns or bugs. Priorities 2-5 choose the newest versions and default variant
values and providers for {\it roots} in the DAG. We prioritize versions first, then
virtual providers (e.g., the user's preferred MPI), then default variant values. After
root configuration, we prioritize non-root configuration (6-7, 11-12). Preferences flow
downward in the DAG via dependency relationships, and if a root demands a particular
version of a dependency, the dependency's preference should not be able to override
this. Ideally, we would enforce strict DAG precedence on preferences (i.e., every node's
preference would dominate those of its dependencies), but we have seen performance
penalties when attempting to track the relative depths of all nodes, so we settle for a
two-level hierarchy of roots and non-roots. We try to enforce consistency in the
resolved graph by minimizing ``mismatches'' (8-9, 14)---these criteria ensure that
neighboring nodes are assigned consistent compilers, OS's and targets unless otherwise
requested. We model preferred compilers, targets, and OS's for non-roots at lower
priority than mismatches, so that dependencies inherit these properties from their
dependents unless set explicitly by the user. All preferred versions, compilers, OS's,
targets, etc. can be overridden through configuration in Spack.

Spack's optimization criteria are unique among dependency solvers in that many of them
are simply not modeled by other systems. In a typical Linux distribution like Debian,
target and toolchain compatibility are not modeled---they are uniform across the DAG. If
we look at {\tt aspcud}~\cite{gebser+:2011-aspcud}, an ASP solver plugin for Debian's
APT, we see that there are around 6 optimization criteria, all to do with
versions and version preferences, far fewer than the 15 criteria modeled here.

\section{Reusing Already Built Packages}
\label{sec:reuse}

While \spack{} is primarily a package manager installing software \emph{from sources},
the ability to reuse already built software and mix it seamlessly with source builds has
been a critical request from the community for several years.

\begin{figure}[t]
  \centering
  \includegraphics[width=\columnwidth]{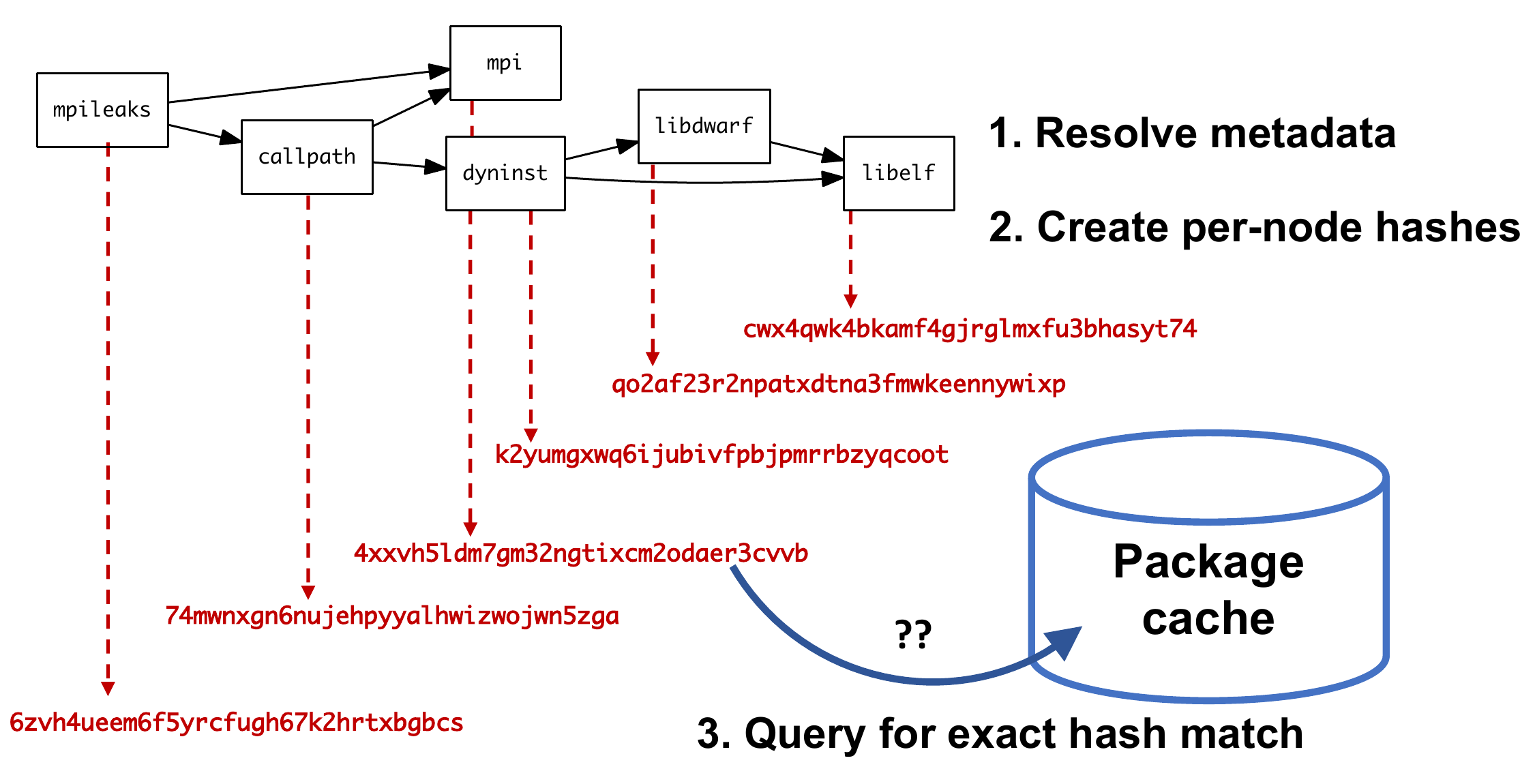}
  \caption{
    Hash-based reuse in the original \spack concretizer.
    \label{fig:hash_reuse}
  }
\end{figure}

Functional packaging systems, including \spack{} with the old concretizer, reuse builds
via metadata hashes. This mechanism relies on the fact that, when an installation graph
is concretized, each node in the DAG is given a unique hash, as shown in
Figure~\ref{fig:hash_reuse}.
Unless the user was explicit, \spack{} only reused packages if their hashes matched
exactly. Small configuration changes could easily result in little or no reuse.



A more effective way to approach software reuse can be achieved by leveraging the
``Generalized Condition Handling'' logic described in
Section~\ref{subsec:generalizedcond}. First, all the metadata from installed packages
can be encoded into facts:
\begin{minted}[fontsize=\footnotesize, bgcolor=bg]{prolog}
installed_hash("zlib","7fatd...").
imposed_constraint("7fatd...","node","zlib").
imposed_constraint("7fatd...","version","zlib","1.2.11").
imposed_constraint("7fatd...","node_platform","zlib","linux").
imposed_constraint("7fatd...","node_os","zlib","ubuntu20.04").
imposed_constraint("7fatd...","node_target","zlib","icelake").
...
\end{minted}
The encoding is based on \texttt{imposed\_constraint}, but the constraint ID is the
hash associated with the installed package. To minimize the number of builds from
source, the solver is allowed to choose a hash to resolve any package:
\begin{minted}[fontsize=\small, bgcolor=bg]{prolog}
{hash(P, Hash) : installed_hash(P, Hash)} 1 :- node(P).
\end{minted}
It imposes all constraints associated with chosen hashes:
\begin{minted}[fontsize=\small, bgcolor=bg]{prolog}
impose(Hash) :- hash(P, Hash).
\end{minted}
And number of builds (packages \emph{without} a hash) is minimized:
\begin{minted}[fontsize=\small, bgcolor=bg]{prolog}
build(P) :- not hash(P, _), node(P).
#minimize { 1@100,P : build(P) }.
\end{minted}

This is a remarkably simple encoding for a complex constraint problem that cannot be
solved with the prior greedy concretizer. However, the devil is in the details.
Minimizing builds can have two effects---it can cause the concretizer to prefer an
existing installation over building a new version of a package, but if set as the
highest optimization priority, it also causes the concretizer to configure newly built
packages in any way that {\it minimizes dependencies}. Generally, though, users expect
new builds to use regular defaults---i.e., most recent version, preferred variants, etc.
As an illustrative example, building \texttt{cmake} with these objectives will build
{\it without} networking capabilities, because it omits \texttt{openssl} and its
dependencies from the graph!



Our solution is to split the optimization criteria into two sets of identical
``buckets'', one for installed software and one for software to be built,
and order their priorities:
\begin{enumerate}
\item Minimize all objectives for software to be built
\item Then, minimize the number of builds
\item Minimize all objectives for already built software
\end{enumerate}

\begin{figure}
  \centering
  \includegraphics[width=.8\columnwidth]{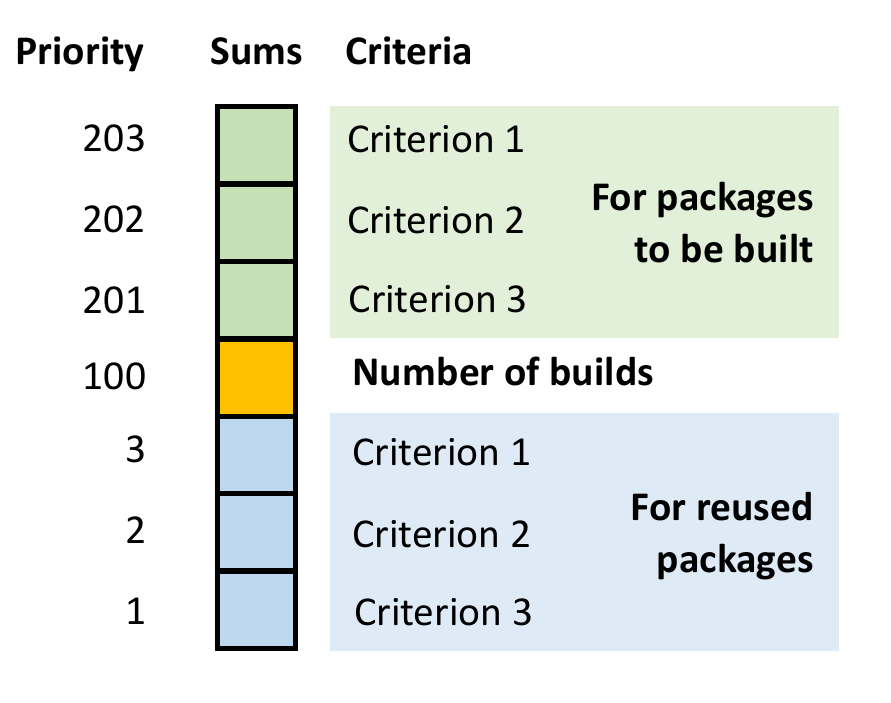}
  \caption{
    Vector of accumulated weights of optimization criteria.
    Contributions to each criterion's objective are divided into two buckets:
    a higher priority bucket if the package is to be built, and a lower priority bucket
    if the package is already installed. Objective vectors are compared lexicographically.
    \vspace{-1em}
    \label{fig:opt-vector}
  }
\end{figure}

\noindent
This is actually fairly simple to encode in ASP:

\begin{minted}[fontsize=\footnotesize, bgcolor=bg]{prolog}
build_priority(P, 200) :- build(P), node(P).
build_priority(P, 0)   :- not build(P), node(P).
% ....
#minimize{
    W@2+Priority,P
    : version_weight(P, W), build_priority(P, Priority)
}.
\end{minted}

For each criterion defined in Table~\ref{table:optimization-criteria}, we include a {\tt
  \#minimize} statement like the one above. It contributes a version weight {\tt W} for
each package {\tt P} in the graph to bucket number {\tt 2 + Priority}. {\tt Priority} is
200 if the package needs to be built, but it is zero if the package is already
installed. The structure of these buckets for a scenario with three criteria is shown in
Figure~\ref{fig:opt-vector}. Buckets are ordered by priority from high to low, and the
three per-criterion buckets for built packages have priorities 203, 202, and 201. In the
middle, at priority 100, is the total number of builds, and below it are all the
original criteria with their original numbers: 3, 2, and 1. This vector is computed for
each model and compared lexicographically to determine the ``best'' one.

If we ensure that all of our optimization criteria are minimizing, i.e., that the
``best'' value for any criterion is zero, then any scenario where a package must be
built will still be worse than one where it is reused. Even if all criteria are
minimized for built packages, the number of builds will still be greater than zero,
making the build scenario worse, because number of builds has higher priority. This
allows \spack{} to pick the best version available from what is installed without
adversely affecting the choice of defaults for built packages. \spack{} will prefer to
build the default configuration of a package that it {\it has} to build, even if that
means building more packages. However, if some installed package {\it can} satisfy a
dependency requirement, \spack{} will prefer to reuse it.

\begin{figure*}[ht]
  \centering
  \subfloat[][With hash-based reuse: all misses]{
    \includegraphics[width=.4\textwidth]{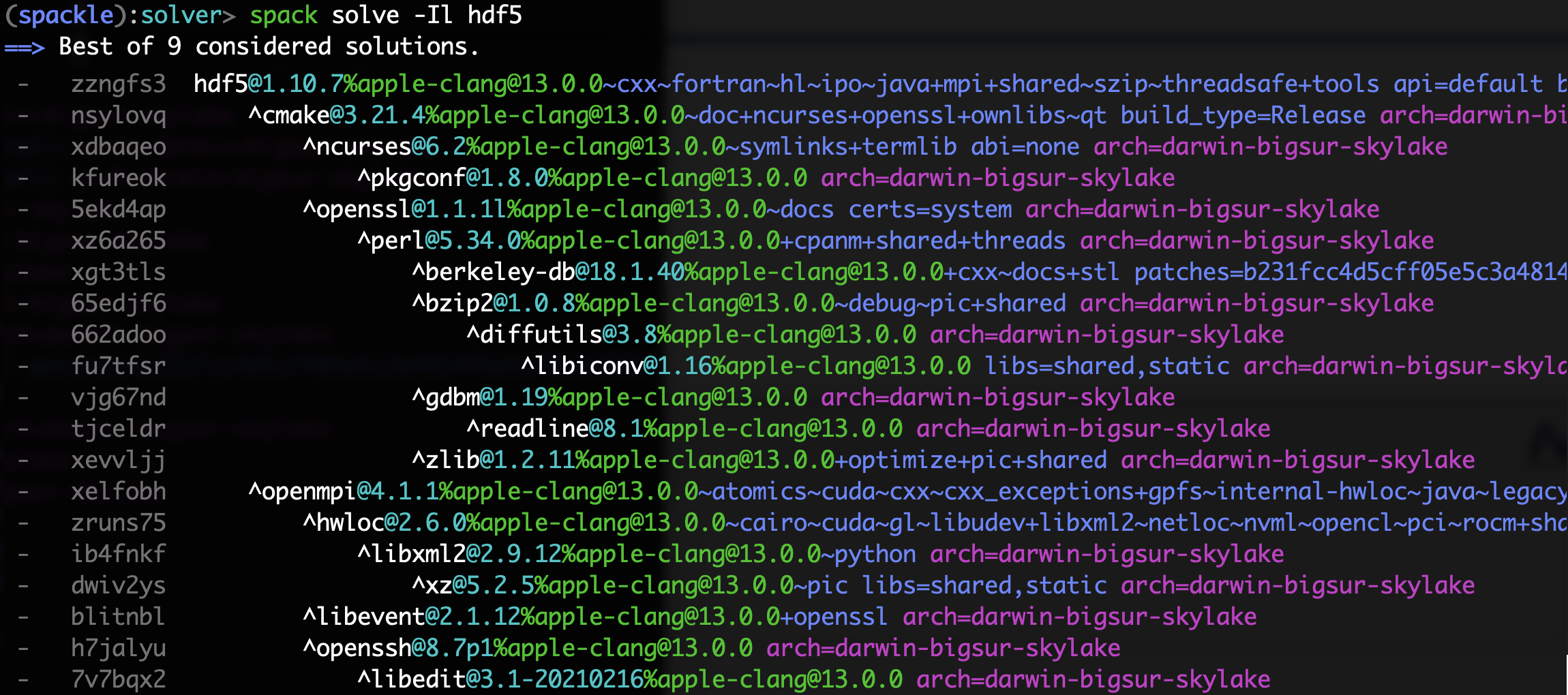}
    \label{fig:noreuse}
  }
  \subfloat[][Solving for reuse: 16 packages reused for {\tt hdf5} build]{
    \includegraphics[width=.43\textwidth]{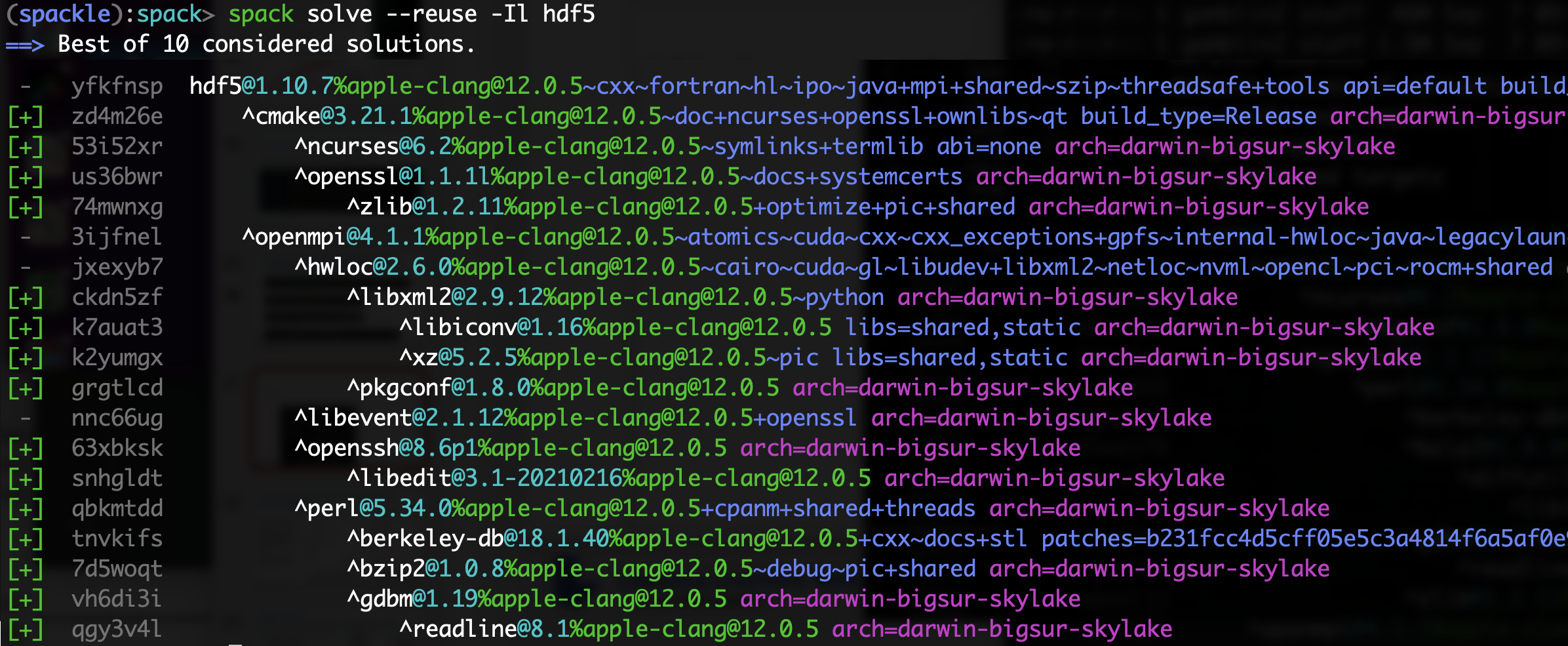}
    \label{fig:reuse}
  }
  \caption{Concretization with and without reuse optimization. \vspace{-1em}}

\end{figure*}

%
%
The benefits of reusing packages are clear. Figure~\ref{fig:noreuse} shows a
concretization relying on purely hash-based software reuse. We can see that no match was
found and 20 installations need to be performed from source. In Figure~\ref{fig:reuse}
we show the same concretization with the reuse logic turned on. In this case 16
installed packages can be reused and only 4 need to be built. As described above, reuse
takes priority over the defaults for already installed software, allowing \spack{} to
reuse \texttt{cmake} 3.21.1 even though the preferred version for a new build is 3.21.4.
This can save users considerable time.

\section{Performance Results}
\label{sec:perf-results}


The \clingo{} solver performance given a logic program depends on a number of factors.
First, the number of facts in a specific concretization. Second, the configuration and
various optimization parameters passed to the solver.

The solving process consists of four stages: \emph{setup}, \emph{load}, \emph{ground},
and \emph{solve}. The first two are preliminary phases and the other two perform the
solve. Specifically, the setup phase generates the facts for the given spec,
whereas the load phase loads the main logic program (i.e., the rules of the
software model) as a resource into the solver. The grounding phase comes first. Once we
have a grounded program, we can run the last phase: the full solve in \clingo{}.
We instrumented the solver to measure each phase.

\begin{figure*}[htb]

    \centering
    \subfloat[][Ground times vs. number of dependencies.]{
    \label{subfig:deps_quartz_load_a}
        \includegraphics[width=\perfsubfigwidth\textwidth]{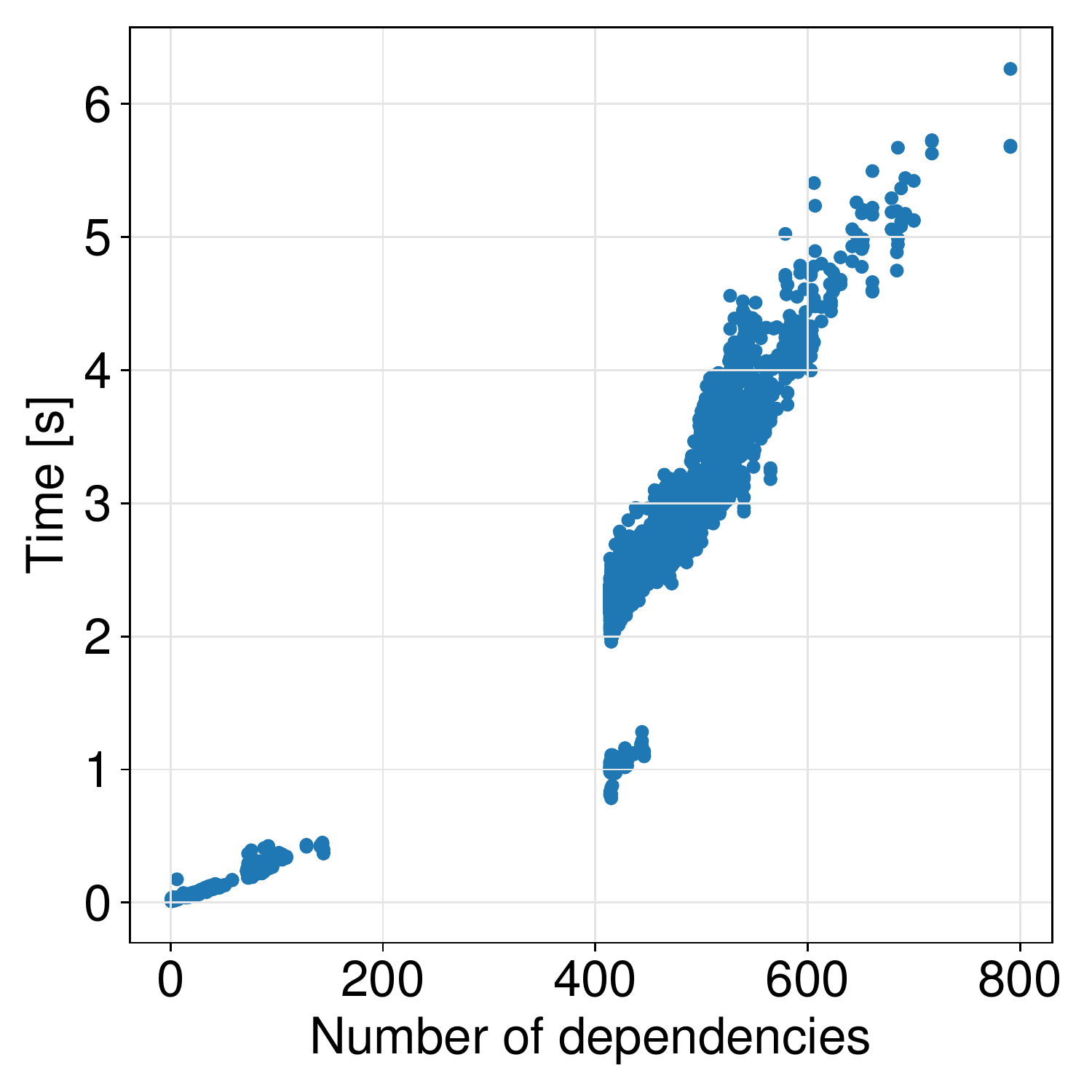}
    }\hfill%
    \subfloat[][Solve times vs. number of dependencies.]{
    \label{subfig:deps_quartz_solve_a}
        \includegraphics[width=\perfsubfigwidth\textwidth]{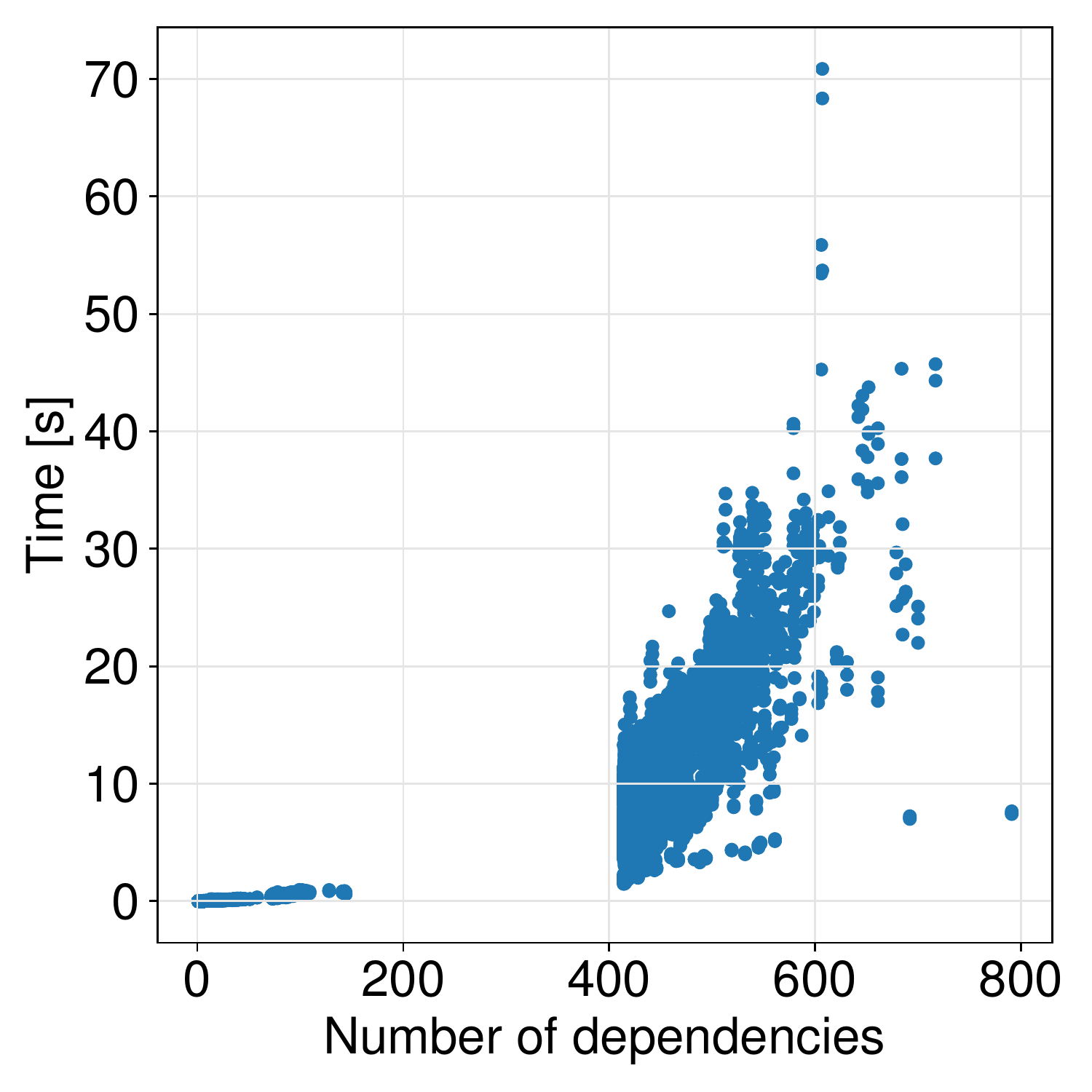}
    }\hfill%
    \subfloat[][Total times vs. number of dependencies.]{
    \label{subfig:deps_quartz_full_a}
        \includegraphics[width=\perfsubfigwidth\textwidth]{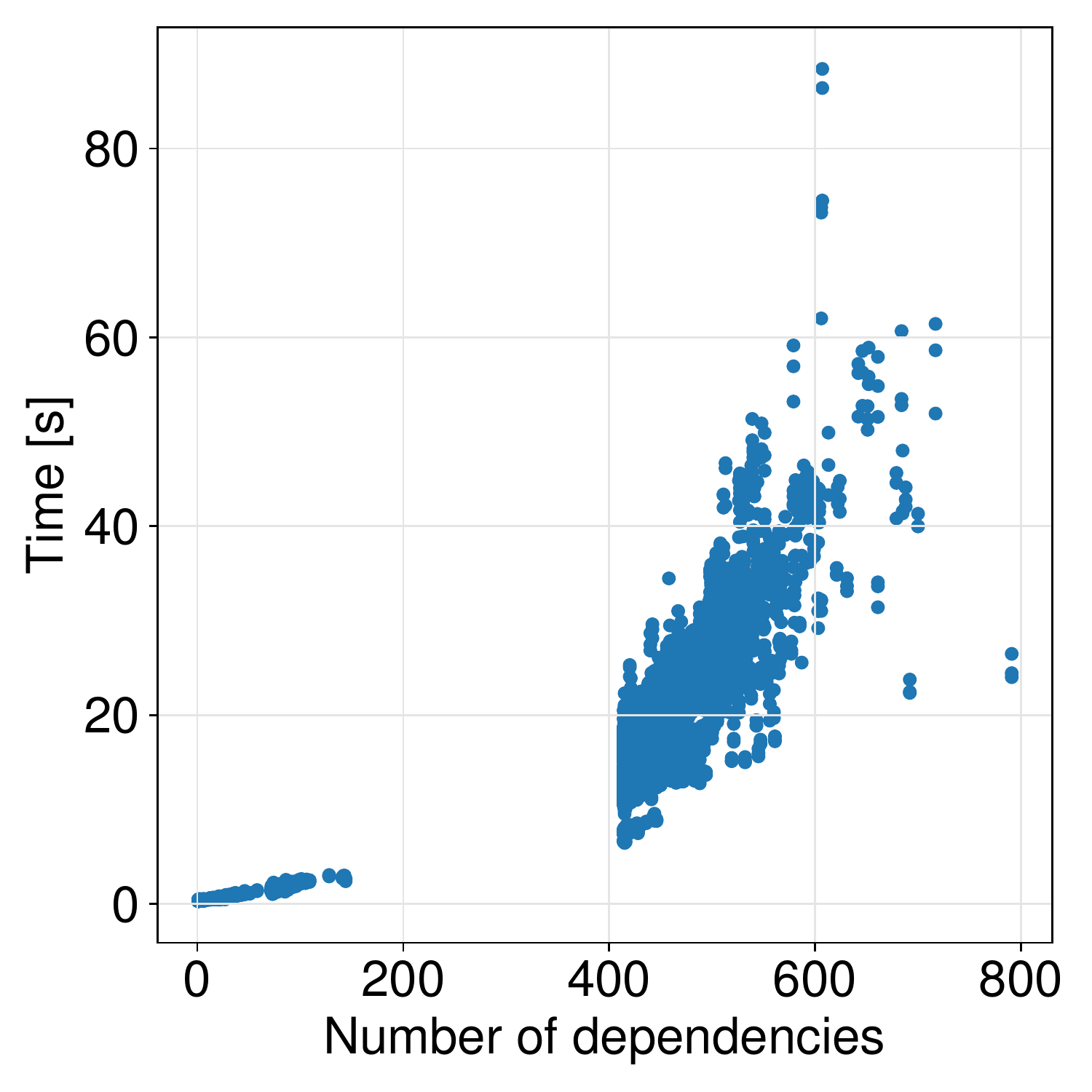}
    }\hfill%
    \subfloat[][CDF of solve time for different {\tt clingo} configurations.]{
    \label{subfig:cdf_quartz_full_a}
        \includegraphics[width=\perfsubfigwidth\textwidth]{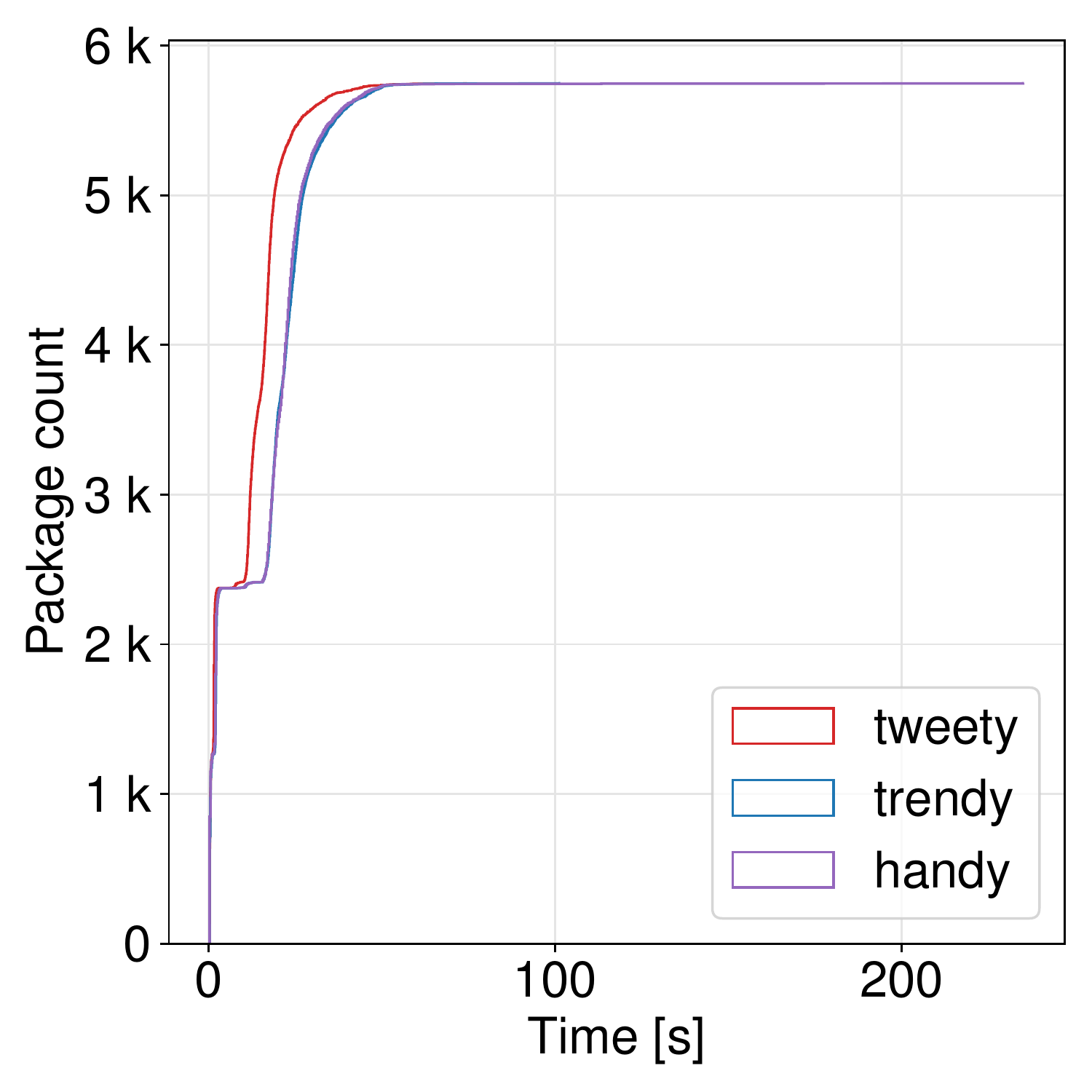}
    }\\
    \subfloat[][CDF of setup times for different cache sizes across all E4S packages.]{
    \label{subfig:cdf_e4s_quartz_load_a}
        \includegraphics[width=\perfsubfigwidth\textwidth]{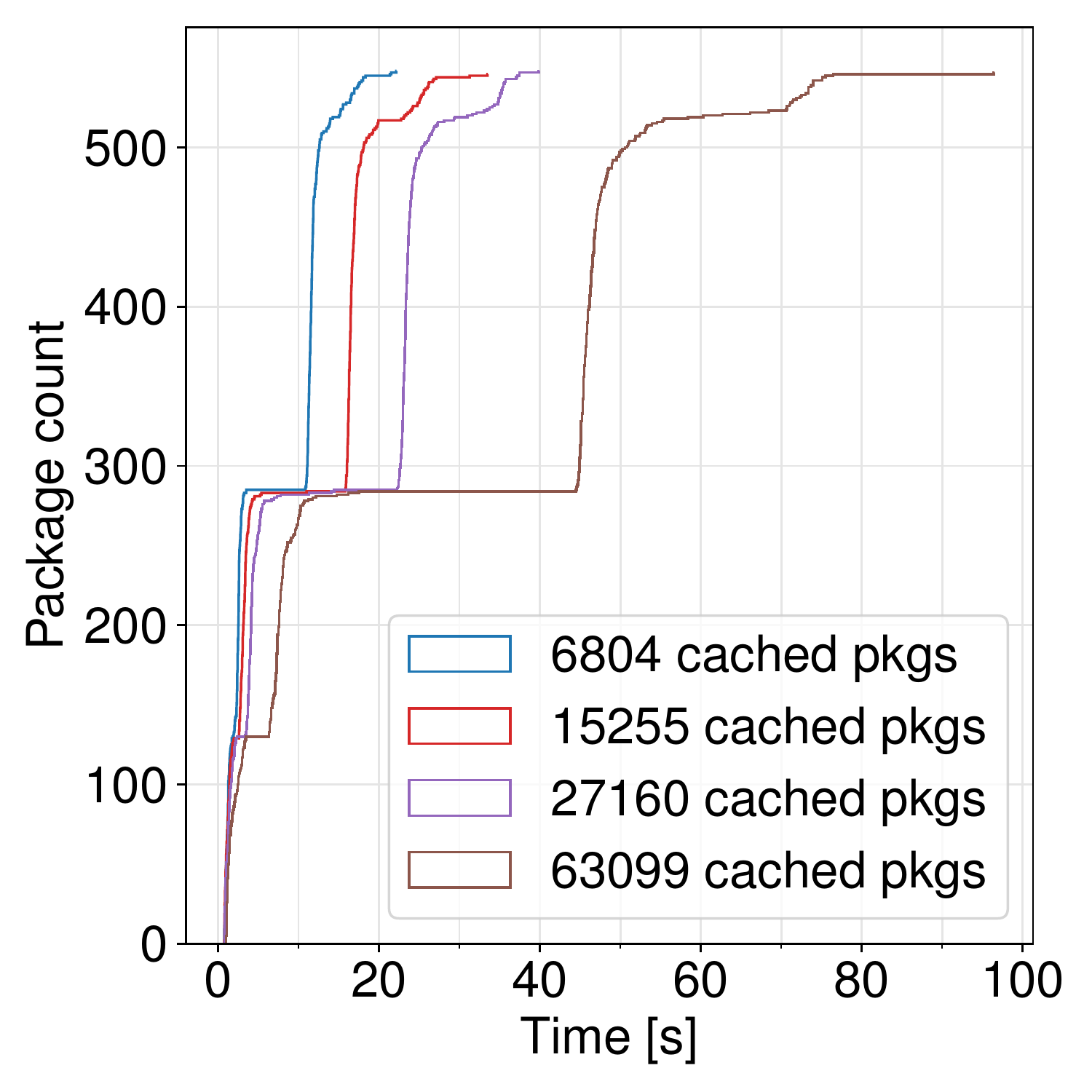}
    }\hfill%
    \subfloat[][CDF of solve times for different cache sizes across all E4S packages.]{
    \label{subfig:cdf_e4s_quartz_solve_a}
        \includegraphics[width=\perfsubfigwidth\textwidth]{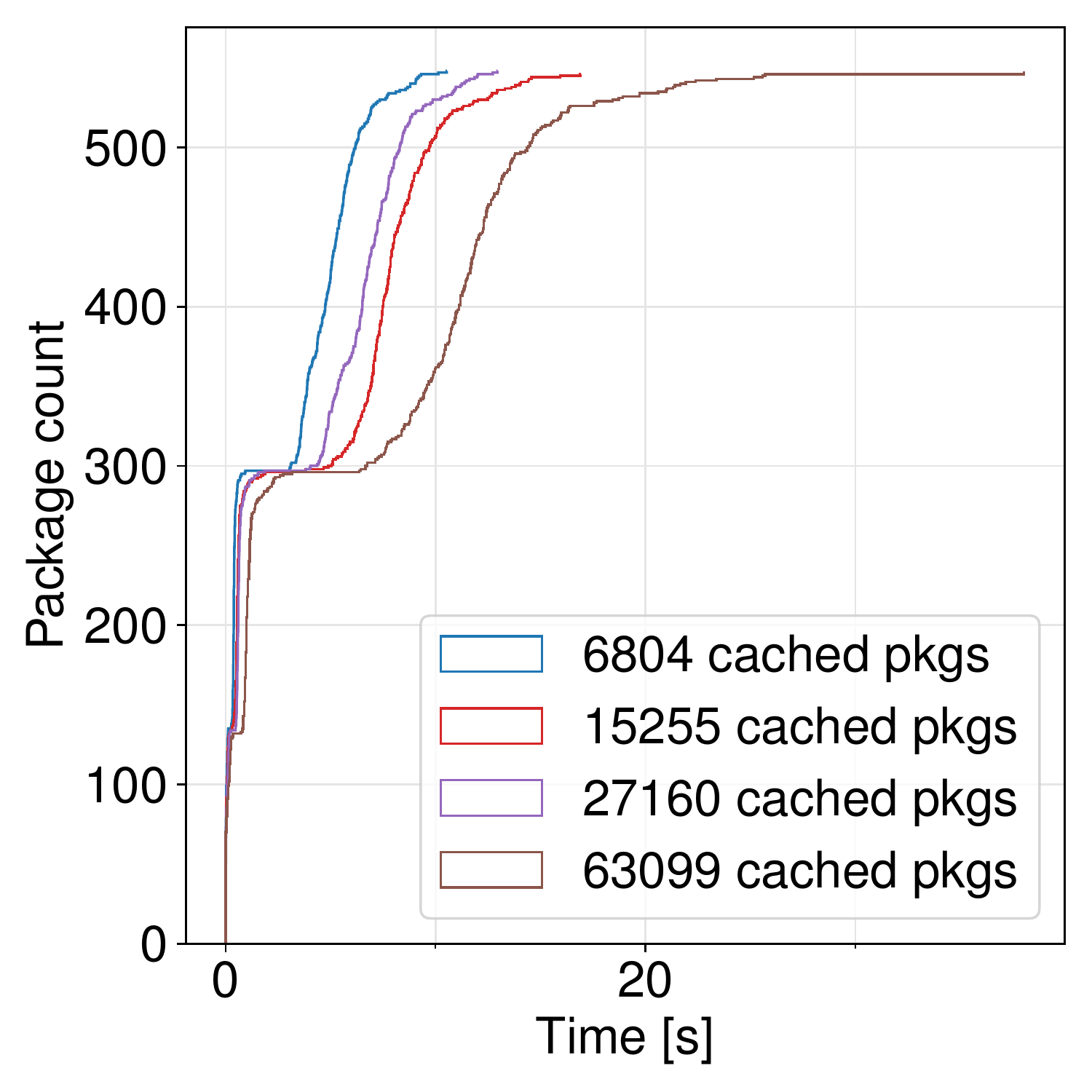}
    }\hfill%
    \subfloat[][CDF of total time for different cache sizes across all E4S packages.]{
    \label{subfig:cdf_e4s_quartz_full_a}
        \includegraphics[width=\perfsubfigwidth\textwidth]{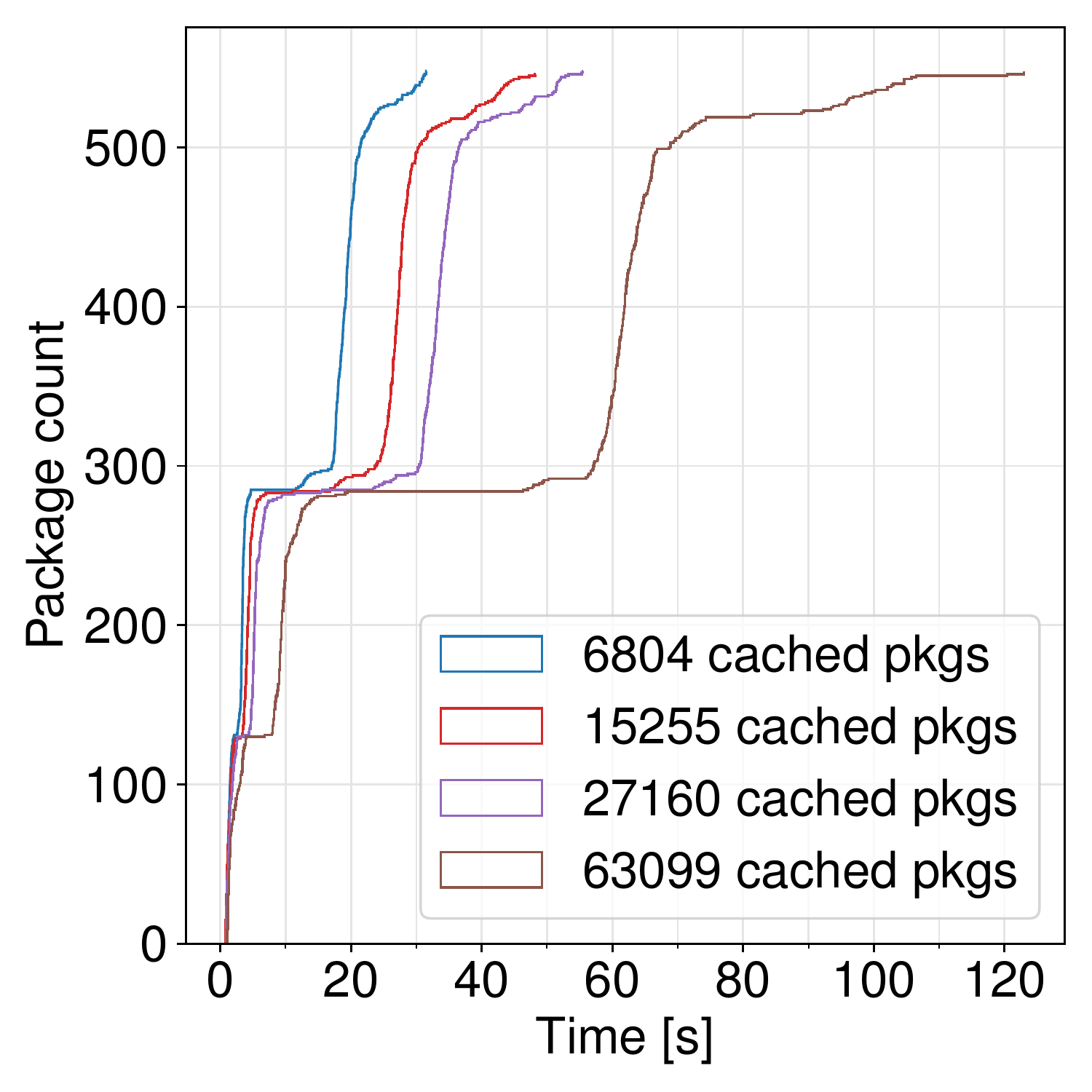}
    }\hfill%
    \subfloat[][CDF of old concretizer times and \clingo{} total solve times.]{
    \label{subfig:cdf_old_vs_new_a}

    \includegraphics[width=\perfsubfigwidth\textwidth]{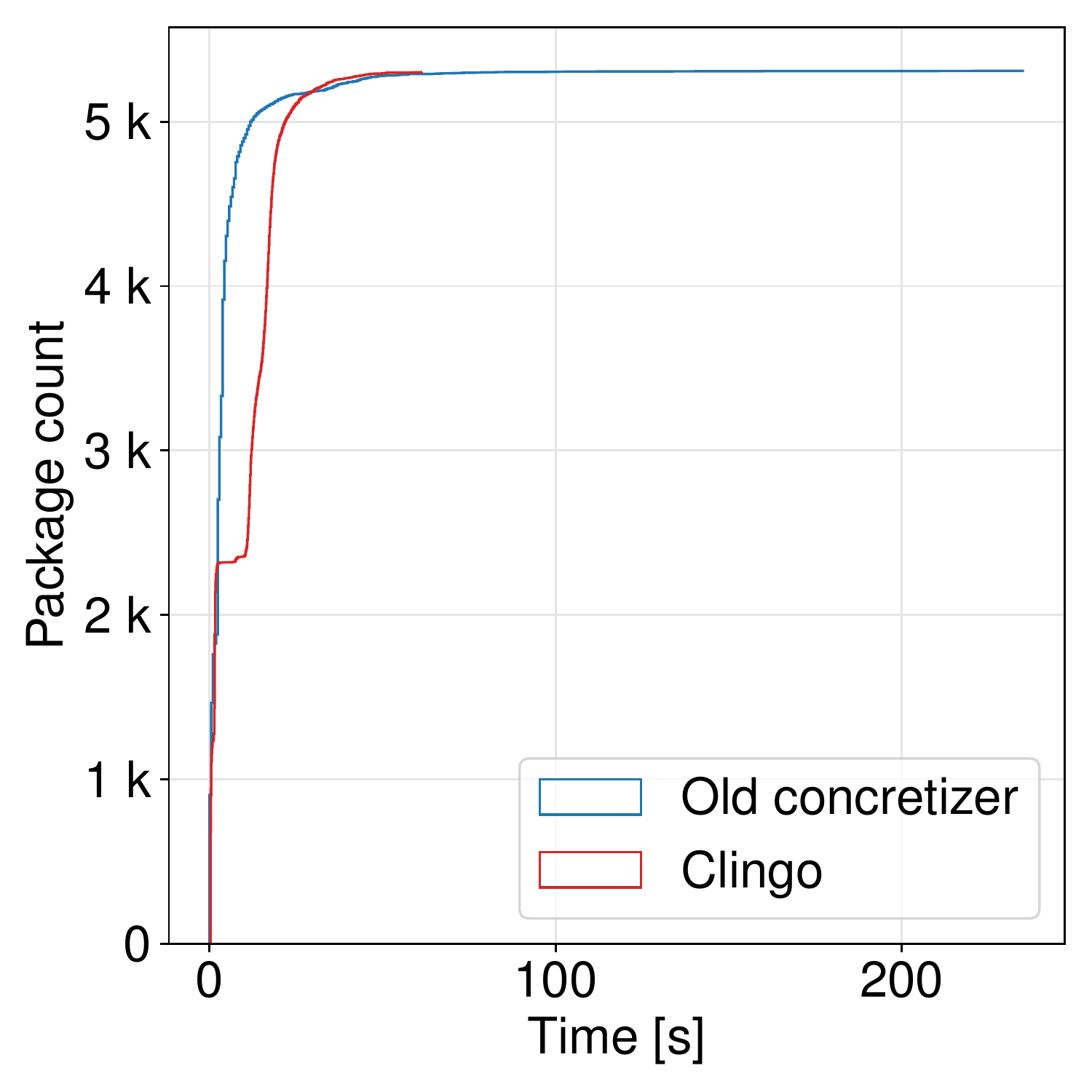}
    }

    \caption{Performance figures of solving times across all packages on Quartz.\vspace{-1em}}
    \label{fig:all_perf}

\end{figure*}

\subsection{System Setup}

All performance runs were executed on a single node each of the Quartz and Lassen
supercomputers at at Lawrence Livermore National Laboratory (LLNL)~\cite{llnl:hpc}.
Quartz is an Intel-based 3.3 PF machine, where each node comprises two Intel Xeon
E5-2695 v4 (Haswell) processors and 128GB of memory. The Lassen machine is a smaller
variant of Sierra, a 125 PF capability supercomputer at LLNL. Each node on Lassen has
two IBM Power9 little-endian processors and four NVIDIA Tesla V100 (Volta) GPUs. There
are 256GB of memory on each node. No hardware accelerators were used in any of our
testing. Experiments ran with NFSv3.

\subsection{Solve timings for all packages}

In this section we examine the solving times for all the packages. First, we focus on
the relation between the solving times and the number of dependencies for each package.

For number of dependencies, we measured the number of possible
dependencies added to the solve, rather than the total dependencies in
the result, because it is a closer measure of the necessary work for
the solver. This leads to natural clustering in the number of
dependencies, as many simple dependencies lead to large numbers of
potential dependencies that dwarf other differences between specs.

Figures~\ref{subfig:deps_quartz_load_a}, \ref{subfig:deps_quartz_solve_a}, and
~\ref{subfig:deps_quartz_full_a} show the grounding, solve, and full solving (i.e.,
involving all the stages) times for all the packages on Quartz. The results on Lassen
are comparable to Quartz and are omitted to save space. Load times, as one would expect,
were not affected by the number of packages. Also, the setup times are the same order of
magnitude as ground times and do not depend on \clingo{}'s performance so they were
omitted in favor of showing times that are directly dependent on \clingo{}. We used the
\clingo{}'s {\it tweety} configuration and unsatisfiable-core-guided optimization strategy
({\it usc,one}) for running the solving process. Further below we explore the differences
in solving times between these different strategies.

We can see from the figures that the time increases as the number of possible
dependencies increases. This is because more dependencies lead
to more facts and a bigger logic program overall. We measure possible
dependencies rather than actual dependencies because possible dependencies
better measure the size of the problem space. In particular, when many packages can
provide a virtual dependency like the Message Passing Interface (MPI), much of the
potential solve space is not present in the final result when a single MPI
implementation is chosen.

Figure~\ref{subfig:deps_quartz_full_a} also shows that there are two major clusters in
the execution times. The clusters are separated by a gap in the possible dependencies.
One cluster contains packages with less than 200 possible dependencies, whereas, the
other major cluster contains packages with more than 400 possible dependencies. This
natural clustering occurs because some low-level dependencies have options that can
trigger huge potential dependency trees, and the gap is between packages that can
include those dependencies and those that cannot. For example, any package that can
depend on MPI in any possible configuration of it and its dependencies involves at least
452 possible dependencies. Since many unusual but technically valid package
configurations can create circular dependencies between MPI and build tools (e.g.
\texttt{mpilander} provides MPI and \texttt{mpilander -> cmake -> qt -> valgrind ->
  mpi}), in practice a gap forms between the packages that (mostly) all can depend on
MPI, and those that cannot. While the solver excludes the configurations that actually
produce circular dependencies, these cycles expand the solution space of the solver and
therefore affect performance.






Besides dependencies another set of factors that influences the execution times are
\clingo{} parameters. Specifically, \clingo{} defines six configuration presets:
\emph{frumpy}, \emph{jumpy}, \emph{tweety}, \emph{trendy}, \emph{crafty}, and
\emph{handy}. Each preset sets numerous low level parameters that control different
aspects of the solver. In our performance study, we specifically focus on three
configurations: \emph{tweety} -- geared towards typical ASP programs, \emph{trendy} --
geared towards industrial problems, and \emph{handy} -- geared towards large problems.

Figure~\ref{subfig:cdf_quartz_full_a} shows the cumulative distribution of full solve
times \emph{tweety}, \emph{trendy}, and \emph{handy} configurations on Quartz. The
results on Lassen are comparable to Quartz and are omitted to save space. The vast
majority of packages are fully solved in under 25 seconds, and most are solved in less
than 10 seconds. We also saw (figure omitted) that there is no difference in ground
times between the different configurations. This suggests that most low level parameters
that are tweaked by each configuration control the actual solving phase. The figures
clearly indicate that \emph{tweety} performs better than the other configurations we
benchmarked. We therefore chose it as our default configuration for concretization.

Figure~\ref{subfig:cdf_old_vs_new_a} shows the the cumulative distribution of the old
concretizer times and \clingo{} total solve times (under the \emph{tweety}
configuration) on Quartz. From it, we see that about 2.2K packages
belong to the cluster with less than 200 dependencies in
Figure~\ref{subfig:deps_quartz_full_a}, which means that the dependency trees of these
packages are smaller. This leads to shorter ground and solve times and that makes
\clingo{}'s times correspond to old concretizer times for these packages. The packages
in the other cluster have potentially huge dependency trees that increase the total
solving times and that is reflected in the deviation of \clingo{}'s times in
Figure~\ref{subfig:cdf_old_vs_new_a} from the old concretizer times.


\subsection{Solve timings for all packages with reuse}

In this subsection, we examine the performance of the solver with the \emph{reuse} flag
switched on. As described in Section~\ref{sec:reuse}, reusing packages in a buildcache
increases the number of facts proportionally to the number of cached packages.

We specifically focus on the packages in the ECP Extreme-scale Scientific Software Stack
(E4S) project~\cite{e4s}. It is a community effort to provide open source software
packages for developing, deploying, and running scientific applications on HPC
platforms. There are just under 600 packages in E4S, but the buildcache of the project
targets different architectures, operating systems, and compilers, thereby totaling over
60K pre-compiled packages (hash signatures). We divided the buildcache into 4 groups:
full buildcache (63099 packages), buildcache restricted to the \texttt{ppc64le}
architecture (27160 packages), buildcache restricted to the \texttt{rhel7} OS (15255
packages), and buildcache restricted to both \texttt{ppc64le} architecture and the
\texttt{rhel7} OS (6804 packages). Benchmarking across an increasing size of the
buildcache provides us with a better understanding of the impact of reuse optimization.

Figures~\ref{subfig:cdf_e4s_quartz_load_a}, \ref{subfig:cdf_e4s_quartz_solve_a},
and~\ref{subfig:cdf_e4s_quartz_full_a} show the cumulative distribution of the solve
times of the E4S packages with increasing buildcache on Quartz. The results on Lassen
are comparable to Quartz and are omitted to save space. Setup times are higher than
solve times, even for smaller buildcaches. This happens because when we reuse packages
we need to load the database of existing packages. This is currently time consuming
because it requires us to read and compare many spec objects in Python. There a jump in
the solve times for the largest buildcache, but most solves take less than 10 seconds.
The fact that the runtime is dominated by serial setup time is good news; setup time is
easily optimized away through caching and optimizing Python code, while solve
time is not. The E4S buildcache, with nearly 64,000 packages, is much larger than most
package repositories, and we expect that our approach will scale to much larger
buildcaches if we can optimize the Python runtime. Multi-shot solver techniques may
offer additional solver performance, as we can divide and conquer for a slightly less
optimal final result.


\section{Conclusions}
\label{sec:conclusions}

The complexity of software dependencies and optimization needs in HPC creates a
particular challenge for package management in the HPC space. Ad-hoc techniques for
dependency resolution have proven to require substantial investment of programmer hours
to manage even a small subset of the possible space of HPC software configurations.

In this paper we introduced a new dependency resolution method for \spack{}, an HPC
package manager. This new dependency resolution method uses Answer Set Programming to
model Spack's DSL, compatibility semantics, and optimization rules in a maintainable,
declarative syntax. It has allowed us to implement new functionality for \spack{}, such
as targeted reuse of installations and binary packages, that was simply infeasible with
previous methods reliant on a greedy fixed-point algorithm.
We showed that the performance of the new concretizer is competitive with
the previous algorithm, and that performance of the reuse capability is scalable.


\section*{Acknowledgments}
This work was performed under the auspices of the U.S. Department of Energy by Lawrence
Livermore National Laboratory under contract DE-AC52-07NA27344. Lawrence Livermore
National Security, LLC ({\tt LLNL-CONF-839332}).

\bibliographystyle{IEEEtranN}
\bibliography{concretizer-paper-sc22}

\end{document}